\documentclass[12pt,a4paper]{article}

\usepackage[british]{babel}

\usepackage[a4paper,top=2cm,bottom=2cm,left=2.5cm,right=2.5cm,marginparwidth=1.75cm]{geometry}

\usepackage[style=apa, backend=biber]{biblatex} 
\DeclareLanguageMapping{british}{british-apa} 
\DeclareFieldFormat[article]{volume}{\apanum{#1}} 

\usepackage{amsmath}
\usepackage{graphicx}
\usepackage[colorlinks=true, allcolors=blue]{hyperref}
\usepackage{hyperref}
\usepackage[title]{appendix}
\usepackage{mathrsfs}
\usepackage{amsfonts}
\usepackage{booktabs} 
\usepackage{caption}  
\usepackage{threeparttable} 
\usepackage{algorithm}
\usepackage{algorithmicx}
\usepackage{algpseudocode}
\usepackage{listings}
\usepackage{enumitem}
\usepackage{chngcntr}
\usepackage{booktabs}
\usepackage{lipsum}
\usepackage{subcaption}
\usepackage{authblk}
\usepackage[T1]{fontenc}    
\usepackage{csquotes}       
\usepackage{diagbox}
\usepackage{nomencl}
\makenomenclature
\usepackage{setspace}

\usepackage{titlesec}
\titleformat{\section} 
  {\normalfont\Large\bfseries}{\thesection.}{1em}{}
  
\usepackage{lineno} 

\rightlinenumbers 

\usepackage{float}   
\usepackage{caption} 


\title{OpenCAEPoro: A Parallel Simulation Framework for Multiphase and Multicomponent Porous Media Flows}

\author[1,*]{Shizhe Li}
\author[1,2]{Chen-Song Zhang}
\affil[1]{\small LSEC, ICMSEC, Academy of Mathematics and System Sciences, Chinese Academy of Sciences, Beijing, China}
\affil[2]{School of Mathematical Sciences, University of Chinese Academy of Sciences, Beijing, China}
\affil[*]{Corresponding author: \texttt{lishizhe@lsec.cc.ac.cn}}

\date{}  

\begin{document}
\maketitle

\begin{abstract}
OpenCAEPoro is a parallel numerical simulation software developed in C++ for simulating multiphase and multicomponent flows in porous media. 
The software utilizes a set of general-purpose compositional model equations, enabling it to handle a diverse range of fluid dynamics, including the black oil model, compositional model, and thermal recovery models.
OpenCAEPoro establishes a unified solving framework that integrates many widely used methods, such as IMPEC, FIM, and AIM. 
This framework allows dynamic collaboration between different methods. 
Specifically, based on this framework, we have developed an adaptively coupled domain decomposition method, which can provide initial solutions for global methods to accelerate the simulation.
The reliability of OpenCAEPoro has been validated through benchmark testing with the SPE comparative solution project.
Furthermore, its robust parallel efficiency has been tested in distributed parallel environments, demonstrating its suitability for large-scale simulation problems.
\end{abstract}


\section{Introduction}
Numerical simulation is crucial for understanding and analyzing flow in porous media across various fields such as hydro geology, petroleum engineering, and environmental science. 
It plays a vital role in informed decision-making, resource optimization, and the sustainable management of natural resources and the environment.
Over the past few decades, emerging technologies in energy and environmental fields, such as shale gas development, geothermal energy extraction, and geological carbon sequestration, have rapidly advanced.
\begin{itemize}
    \item Shale gas is an important source of clean energy, and the technique of hydraulic fracturing has significantly boosted natural gas production. However, this extraction method can lead to environmental issues such as water contamination and methane leakage, necessitating meticulous management~\parencite{vidic_impact_2013,vengosh_critical_2014}. Meanwhile, the geological complexity of shale formations, characterized by low permeability, heterogeneity, natural fractures, and hydraulic fractures, presents significant challenges for numerical simulation~\parencite{zhang_flow_2019}.

    \item Geothermal energy is a renewable energy source that harnesses the natural heat of underground rocks and fluids for power generation and heating. Typically, this process involves drilling in high-temperature rock layers or reservoirs. However, long-term extraction may pose risks such as induced seismicity, groundwater contamination, and thermal pollution. Geothermal simulation requires coupling multiple physical processes, including fluid flow, heat conduction, chemical reactions, and rock mechanics, adding further complexity to numerical modeling~\parencite{blocher_3d_2010}.

    \item Carbon capture and storage (CCS) is a technology that captures carbon dioxide (CO\textsubscript{2}) from emission sources and injects it into underground reservoirs for long-term storage. Various methods have been developed to enhance the efficiency and effectiveness of CCS. For instance, Mineral carbonation reacts CO\textsubscript{2} with minerals like olivine to form stable carbonates, providing permanent storage and utilizing abundant mineral resources~\parencite{wang_kinetics_2019}. CCS can also be combined with enhanced gas recovery from shale reservoirs, where CO\textsubscript{2} injection increases gas recovery and stores CO\textsubscript{2} underground~\parencite{mohagheghian_co2_2019}, yielding a dual benefit. Due to the geological complexity of reservoirs, spatial variations in porosity and permeability, and the complex interactions between CO\textsubscript{2}, reservoir fluids, and rocks, numerical models must accurately capture these interrelated processes.
\end{itemize}

Numerical simulation is crucial for assessing the feasibility and performance of these engineering challenges, optimizing reservoir enhancement techniques, and mitigating potential environmental risks~\parencite{chen_computational_2006}. 
Overcoming these challenges requires interdisciplinary collaboration, advanced modeling and solution technologies, high-performance computing, and continuous validation with field data.
Numerous methods have been proposed in the field of reservoir numerical simulation.
These include Implicit Pressure Explicit Concentration (IMPEC) and Implicit Pressure Explicit Saturation (IMPES)~\parencite{acs_general_1985,watts_compositional_1986}, which offer rapid solutions per time step but are limited by the Courant-Friedrichs-Lewy(CFL) condition~\parencite{coats_impes_2001}, and the Fully Implicit Method (FIM)~\parencite{coats_equation_1980}, which is robust and allows for larger time steps but is computationally expensive. 
The Adaptive Implicit Method (AIM)~\parencite{thomas_reservoir_1983,collins_efficient_1992} strikes a middle ground between these two. 
Moreover, a class of Sequential Fully Implicit (SFI) methods~\parencite{jenny_adaptive_2006,jiang_nonlinear_2019} sequentially solves nonlinear equations in batches using implicit techniques, providing a balance between robustness and computational efficiency.

On the other hand, rapid advancements in computer hardware have provided powerful computational capabilities for increasingly complex reservoir numerical simulations. The prevalence of high-performance computing (HPC) clusters, graphics processing units (GPUs), and parallel computing architectures enables researchers to simulate larger-scale models, incorporate more physical processes, and address complex problems within reasonable time frames. This progress has laid a solid foundation for developing more detailed and realistic simulation models and has driven innovation and breakthroughs in reservoir engineering~\parencite{esler_realizing_2014,liu_performance_2016}. 
However, this also poses new challenges for software design---Software must have a flexible architecture to meet continuously changing simulation needs. 
Developers should employ advanced parallelization techniques, optimize data management and communication strategies, and introduce scalable algorithms so that simulation systems can adapt to the diversity and complexity of the hardware, resulting in more efficient and accurate reservoir numerical simulations~\parencite{hayder_challenges_2012,bogachev_high-performance_2018}.

In this paper, we present OpenCAEPoro\footnote{\url{https://github.com/OpenCAEPlus/OpenCAEPoroX}}, an open-source and cross-platform parallel simulation framework. 
Designed for studying multiphase and multicomponent flows in porous media, OpenCAEPoro aims to integrate diverse models and solution methods into a unified framework, allowing users to implement desired extensions efficiently. 
OpenCAEPoro is characterized by the following features:
\begin{enumerate}
	\item[(1)] We utilize the general compositional model equations and modularize the computations of various physical objects and processes, streamlining the integration of new models and methods.
	\item[(2)] We have developed a unified solution framework that accommodates various methods, such as the widely used FIM, IMPEC, and AIM. This framework enables adaptive switching between these methods at both the nonlinear step and time step levels. This enhancement not only increases the software's flexibility but also promotes collaboration among different methodologies.
	\item[(3)] We have developed a general method based on the domain decomposition concept, referred to as adaptively coupled domain decomposition methods. This approach can adaptively identify significant couplings between subdomains and solve these subdomains in a locally coupled manner during the simulation, thereby accelerating the convergence of nonlinear solvers while maintaining high parallel efficiency. Additionally, this method can provide initial solutions for global methods during the nonlinear solving iterations, with experimental results demonstrating its potential to accelerate large-scale parallel simulation processes.
\end{enumerate}

The subsequent sections are organized as follows: 
Section 2 introduces the mathematical framework employed in the program. 
Section 3 focuses on the software design, including the overall architecture, design principles, and parallelization implementation. 
Section 4 presents the adaptively coupled domain decomposition method.
Section 5 showcases some numerical experiments. 
The final section provides a summary.

\section{Mathematical Equations and Solution Workflow}\label{sec:Mathematical Equations and Solution Workflow}
\subsection{Mathematical Equations}\label{subsec:Mathematical Equations}
We describe the multiphase and multicomponent flow in porous media using the component material conservation equations~\parencite{chen_computational_2006}.
    \begin{equation} \label{eq:mass-conservation}
		\frac\partial{\partial t}{N_{i}}+\nabla\cdot\sum_{j=1}^{n_p}\left(x_{ij}\xi_{j}\mathbf{u}_j-\xi_{j}\mathbf{D}_{ij}\nabla(x_{ij})\right)=Q_i,\quad i=1:n_c
	\end{equation}
	Generally, based on Darcy's Law, we have
	\begin{equation} \label{eq:darcy-law}
		\mathbf{u}_j=-\frac{\boldsymbol{\kappa}\kappa_{rj}}{\mu_j}(\nabla P_j-\rho_{j}g\nabla z),\quad j=1:n_p
    \end{equation}
For non-isothermal conditions, we need to incorporate the energy conservation equation.
    \begin{equation} \label{eq:energy-conservation}
		\frac\partial{\partial t}\left(\phi\sum_{j=1}^{n_p}\xi_jS_jU_j+(1-\phi)U_r\right)+\nabla\cdot\sum_{j=1}^{n_p}\xi_j\mathbf{u}_jH_j-\nabla\cdot(\boldsymbol{\kappa}_T\nabla T)=\sum_{j=1}^{n_p}q_{H,j}-q_{\text{loss}}.
    \end{equation}	
Additionally, some constraints should be added among these physical quantities.
    \begin{align}	
		N_{i} = \phi\sum_{j=1}^{n_{p}}S_{j}\xi_{j}x_{ij},&\quad i=1:n_{c} \\
		\sum_{i=1}^{n_c}x_{ij} = 1,&\quad j=1:n_{p} \\
		\sum_{j=1}^{n_p}S_j = 1,&\\
		V_{f} = V_{p},&
    \end{align}
    
Depending on the model equations employed, the relationships among physical variables differ. 
For example, in the classic black oil model, phase saturation $S_{j}$, the molar fraction of components in phases $x_{ij}$, and molar density $\xi_{j}$ can be explicitly obtained from pressure $P$ and the molar density of components $N_{i}$. 
In contrast, for compositional models, the complex phase equilibrium calculations are necessary. 
Besides, fluid permeability properties, rock compressibility properties, convection-diffusion characteristics of phases and components, as well as well behaviors, can be described by various models. 
Nevertheless, these relationships can remain hidden and need not be explicitly represented in the conservation equations, thereby facilitating their abstraction and modularization.

\subsection{Simulation workflow}\label{subsec:Top-level workflow}

\begin{figure}[H]
    \centering
    \includegraphics[width=0.9\linewidth]{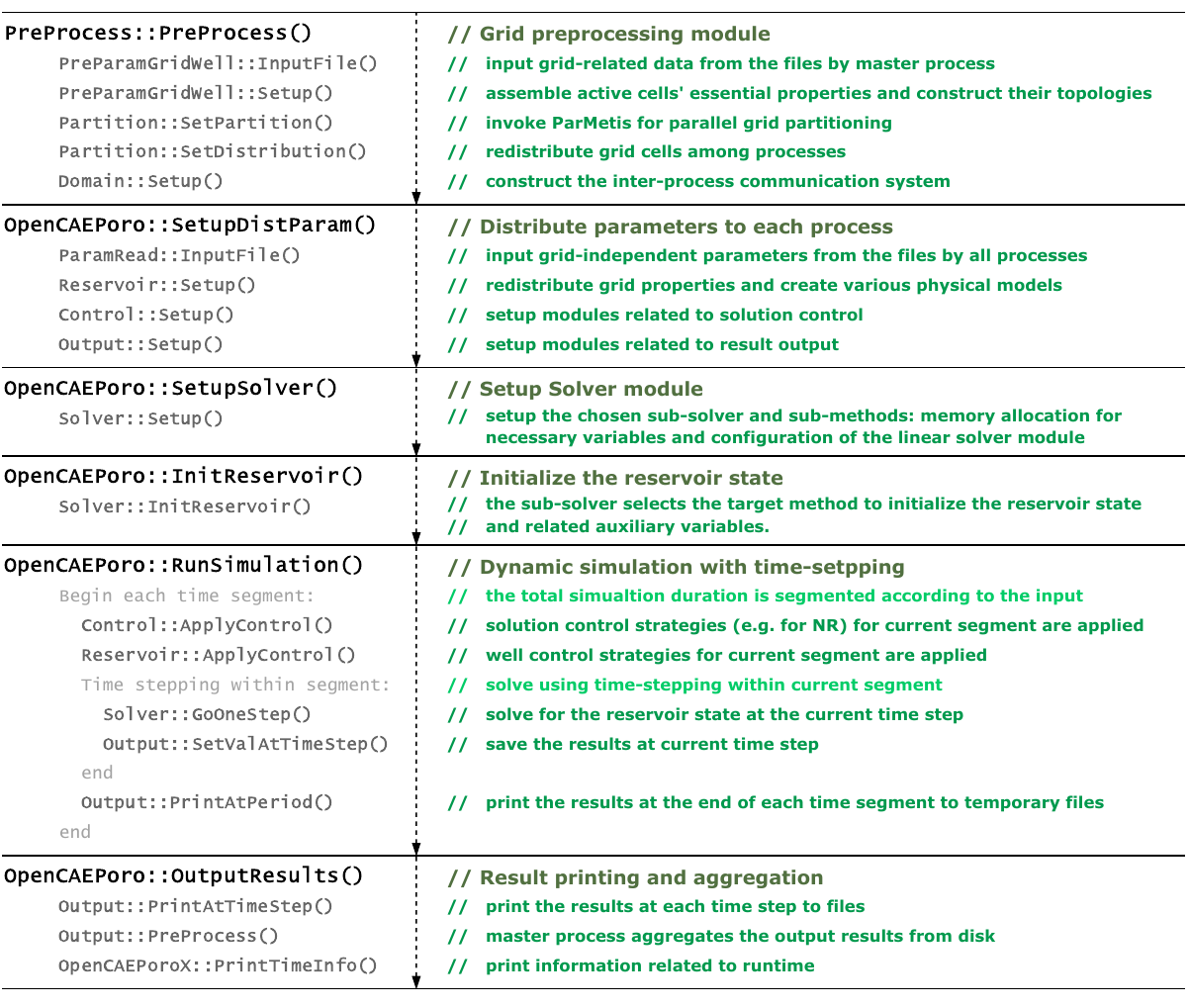}
    \caption{The simulation workflow of OpenCAEPoro}
    \label{fig:totalflow}
\end{figure}

For the time being, in OpenCAEPoro, we have integrated several solution methods including IMPEC, FIM, AIM, and FIMddm (FIM based on domain decomposition). 
These methods are capable of solving various model problems such as black oil models, compositional models, and thermal recovery models.
Figure~\ref{fig:totalflow} illustrates the top-level simulation workflow of OpenCAEPoro, divided into six principal blocks. These blocks handle the preprocessing of the grid, distribution of parameters, setup of solution methods, computation of the initial reservoir state, dynamic simulation, and result output. 
The most critical components of each block are identified and detailed with annotations. 

We provide a few remarks on this workflow depicted in Figure~\ref{fig:totalflow}.
\begin{itemize}
    \item Within \texttt{OpenCAEPoro::RunSimulation}, the overall simulation duration is segmented for two reasons. For practical engineering purposes, conditions like well parameters need frequent adjustments (e.g., different control strategies at different developing stages), and it's essential to monitor and record the reservoir state at critical time points. In contrast, from a numerical simulation perspective, as the simulation progresses, significant shifts in the reservoir state and boundary conditions affect the nature and complexity of the underlying nonlinear and linear problems. This requires varied solution strategies, such as adjusting parameters for Newton-Raphson (NR) and linear solver, choosing suitable time steps, even employing alternative solution methods.
    \item During the parameter file input and result file output stages, a hybrid serial-parallel approach is adopted. In the stage of input, the master process reads grid-dependent parameters, while all processes load grid-independent parameters. This method effectively prevents the difficulties of parallel reading of complexly formatted grid-related data and enhances compatibility with various grid types. 
    While in the output stage, we opted not to consolidate results via inter-process communication for centralized output by the master process. Instead, each process independently writes its portion directly to disk. Once the simulation is completed, the master process gathers and organizes these outputs. This strategy effectively mitigates high memory usage and extensive data communication during the simulation. It also decouples the post-processing of result files from the simulation itself, enhancing flexibility and capacity in managing outputs.
\end{itemize}

\section{Software Design}\label{sec:Software Design}
\subsection{Top-level structure}\label{subsec:Top-level structure}
\begin{figure}[htbp]
    \centering
    \includegraphics[width=0.9\linewidth]{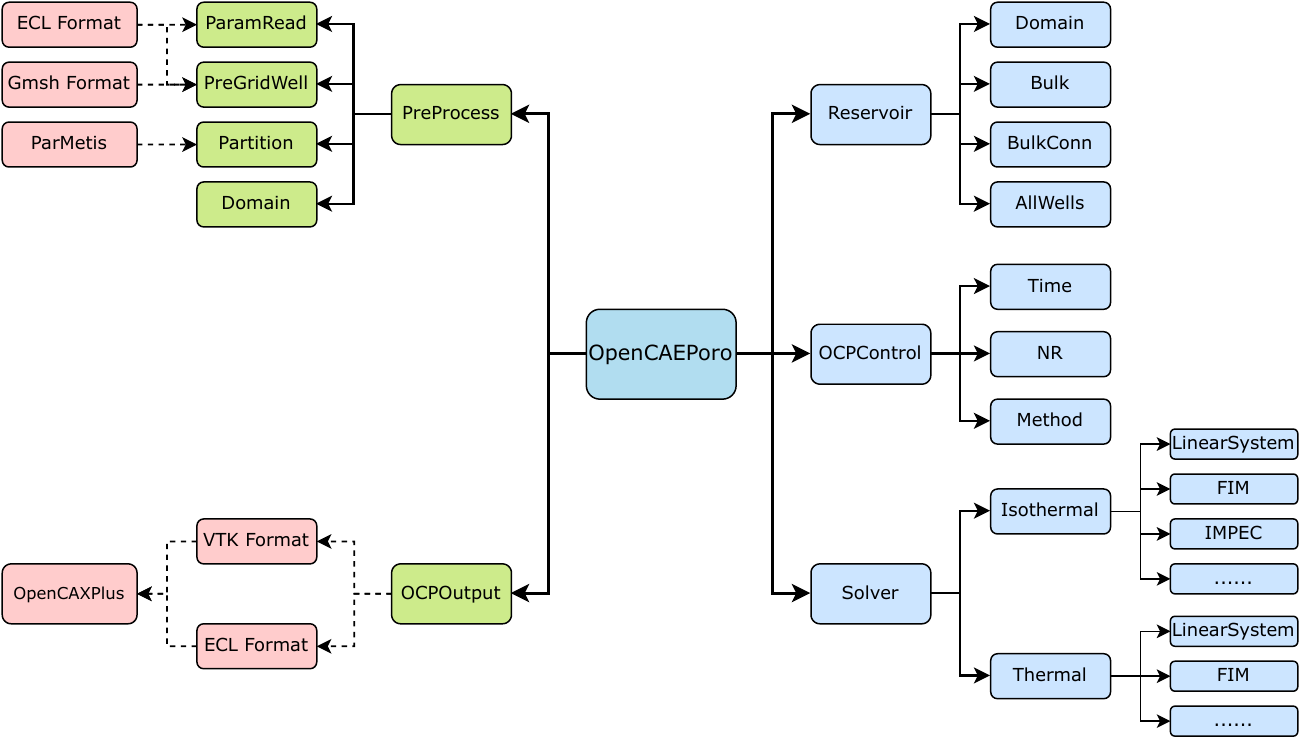}
    \caption{The top-level structure of OpenCAEPoro.}
    \label{fig:TopStructure}
\end{figure}
OpenCAEPoro is constructed based on object-oriented modular principles and contains five major components at the top level:
\begin{enumerate}
	\item \textbf{\texttt{PreProcess} module}: It is responsible for inputting discrete grids and reservoir parameters, as well as performing parallel grid partitioning and redistribution of reservoir information. OpenCAEPoro supports reading input files with ECL-style keywords and unstructured grids in the Gmsh format~\parencite{geuzaine_gmsh_2009}. Additionally, ParMetis~\parencite{karypis_parmetis_1997} is employed for grid partitioning when necessary.
	\item \textbf{\texttt{Reservoir} module}: It contains physical information related to the reservoir, which can be categorized into three parts: information specific to individual grid cells (e.g. PVT calculations), interactions between grid cells (e.g. flux calculations), and interactions between grid cells and wells (e.g. well models). They correspond to the \texttt{Bulk} module, the \texttt{BulkConn} module, and the \texttt{AllWells} module, respectively. Additionally, the \texttt{Domain} module stores information needed by inter-process communications.
    \item \textbf{\texttt{Solver} module}: It accommodates solvers for isothermal and non-isothermal processes. A variety of solution methods, such as IMPEC and FIM, are implemented. The \texttt{Solver} module provides a cohesive interface for the these solvers, which, in turn, establish a unified framework for all underlying solution methods. The \texttt{LinearSystem} module, which includes internal matrix structures and linear solvers, functions at the same level as these methods and is shared among them. 
	\item \textbf{\texttt{OCPControl} module}: It manages the simulation process through two types of controls: operational control and solution control. Operational control focuses on aspects independent of the solution process, such as maximal runtime duration; Solution control encompasses the prediction of time steps, management of nonlinear iterations, as well as the selection of solution methods. 
	\item \textbf{\texttt{OCPOutput} module}: It prints the designated results for subsequent visualization and data analysis. These results are generated at each time step or according to a specified time interval. Currently, OpenCAEPoro supports outputting result files in ECL style and VTK format~\parencite{schroeder_visualization_2006}. 
\end{enumerate}

The \texttt{Reservoir} module and the \texttt{Solver} module are critical in OpenCAEPoro. 
The \texttt{Reservoir} module encompasses all physical variables, objects, models, and their computations related to the reservoir. Meanwhile, the \texttt{Solver} module is responsible for organizing these elements to construct the desired solution workflow. 
The details of their design will be explored in subsequent subsections.

\subsection{\texttt{Reservoir} module}\label{subsec:Reservoir}
The \texttt{Reservoir} module is the core component of OpenCAEPoro. It stores all reservoir information, categorized into three primary classes based on their applicability: \texttt{Bulk}, \texttt{BulkConn}, and \texttt{AllWells}. 
In the context of parallel computing, an additional sub-module, \texttt{Domain}, is required to manage grid partitioning and inter-process communication for message passing.

\subsubsection{Data structures}\label{subsubsec:Data structures}
In OpenCAEPoro, data is stored using the Structure of Arrays (SOA) format, offering several advantages over the Array of Structures (AOS) data structure. 
SOA supports diverse memory access patterns efficiently, which enhances programming flexibility. 
For example, during various stages of the solution process, it is often necessary to access a specific (small) subset of variables from grid cells. 
In such cases, the AOS format could impede efficient memory access. Moreover, the variables required for storage will vary depending on the models and solution methods employed, the AOS format can lead to unused variables occupying memory space, thus creating unnecessary overhead of storage and resulting in substantial cache-misses.
\begin{figure}[H]
		\centering
		\includegraphics[width=0.9\linewidth]{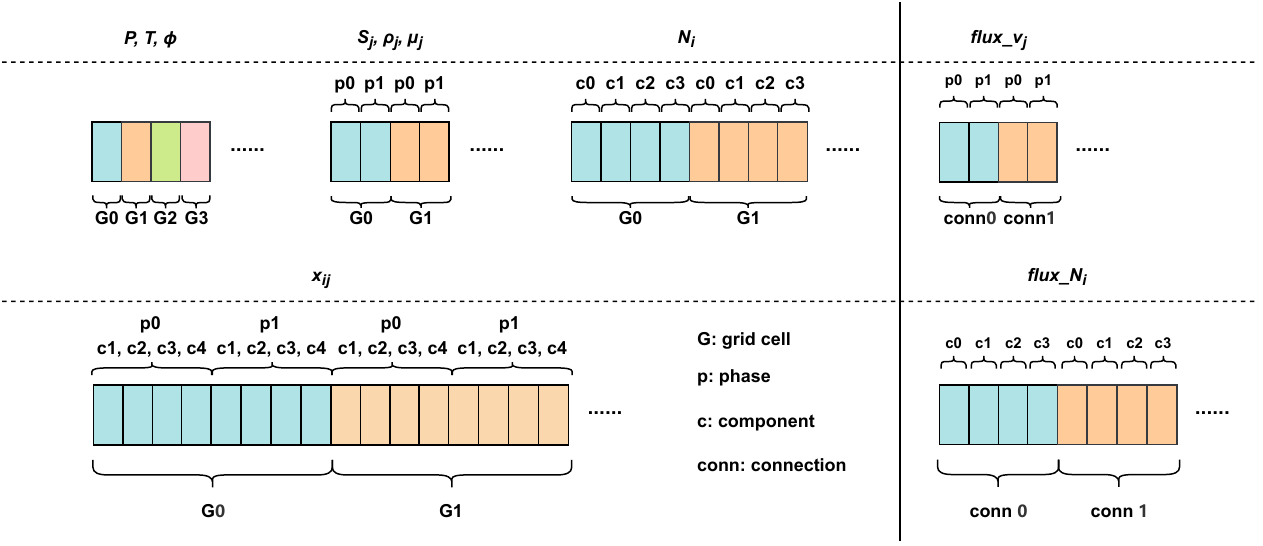}
		\caption{The Structure of Arrays (SOA) data structure is utilized in OpenCAEPoro, illustrated with examples where the number of phases and components are 2 and 4, respectively.}
	\label{fig:OCP-SOA}
\end{figure}

The variables associated with grid cells are stored sequentially according to the order of components, phases, and grid cells. Similarly, the variables related to grid connection pairs are organized by the order of components, phases, and connection pairs. See Figure~\ref{fig:OCP-SOA} for a demo.
Data is exclusively stored for active grid cells. For example, in isothermal flow simulations, grid cells that will never contain movable fluid need not be computed. Consequently, these inactive cells are excluded and their information is not stored.

\subsubsection{Modular design}\label{subsubsec:Modular design}
\begin{figure}[H]
		\centering
		\includegraphics[width=0.3\linewidth]{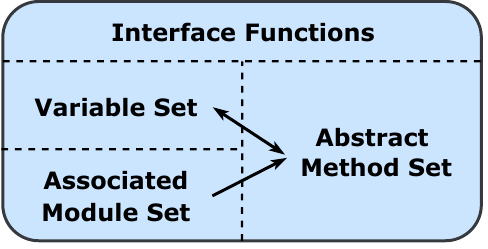}
		\caption{The typical style of small modules in OpenCAEPoro.}
		\label{fig:OCP-module}
    \end{figure}

OpenCAEPoro is partitioned into various modules. Small modules generally are designed for very specific physical objects or processes and own a consistent design language. These small modules are stored within larger modules, either as pointers or entities. Figure~\ref{fig:OCP-module} illustrates the typical style of small modules in OpenCAEPoro, which generally comprising four components:
\begin{enumerate}
	\item \textbf{Interface Functions}: Explicitly defines the responsibilities of the module and specifies its interactions with external systems. Additionally, the module needs to provide external interfaces to configure its internal components, including allocating and initializing required variables, linking to necessary modules, and setting up methods for use.
	\item \textbf{Variable Set}: Required parameters or computed results associated with grid cells or connection units, as determined by the methods in the \texttt{Abstract Method Set}.
	\item \textbf{Associated Module Set}: The set of modules that the current module depends on, i.e. certain methods in the current module needs the data or methods from these associated modules. These modules can be stored in two ways: entity-based (exclusive) and pointer-based (shared).
	\item \textbf{Abstract Method Set}: Usually encompasses a variety of models for implementing the current module. These methods are invoked through a unified abstract interface, leveraging polymorphism. In principal, these methods operate on an individual grid cell or connection pair and the methods that will be used in a simulation run are stored in an array of pointers.
\end{enumerate}
    
\begin{figure}[H]
	\centering
	\includegraphics[width=0.8\linewidth]{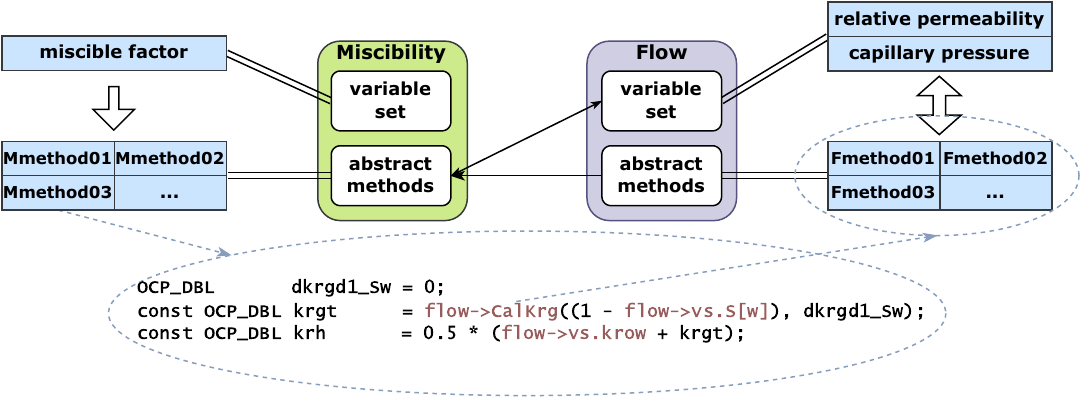}
	\caption{Advantages of method abstraction design in module collaboration}
	\label{fig:OCP-module01}
\end{figure}

In reservoir engineering, numerous models exist for various physical objects and processes, each with distinct characteristics suited to specific scenarios. 
Furthermore, these models can be interrelated (e.g., used together).
Figure~\ref{fig:OCP-module01} presents an example. The \texttt{Flow} module computes fluid permeability properties, such as relative permeability and capillary pressure, which are stored in its variable set. 
It offers several available methods, each representing different models. 
Meanwhile, the \texttt{Miscibility} module focuses on the miscibility properties between phases, which influence phase permeability in multiphase flow.
The code snippet in Figure~\ref{fig:OCP-module01} presents the implementation of \texttt{Mmethod03} within the \texttt{Miscibility} module, which uses the miscible factor to correct the phase permeability. 
Notably, its computational process simultaneously involves the variable set (\texttt{flow->vs}) and computational method (\texttt{flow->Calkrg()}) from the \texttt{Flow} module.
By abstracting methods, we can eliminate the need to be concerned with the internal implementation of \texttt{flow->Calkrg()}. Consequently, it is unnecessary to create a specialized version of \texttt{Mmethod03} for each \texttt{Fmethod} in the \texttt{Flow} module.
We only need to focus on the compatibility between these methods.
By integrating model methods describing the same physical object or process within a unified framework, we can foster collaboration among different models, significantly reduce code redundancy, enhance program maintainability, and improve readability.

\subsubsection{\texttt{Reservoir::Bulk} module}\label{subsubsec:bulk}
\begin{figure}[H]
	\centering
	\includegraphics[width=0.9\linewidth]{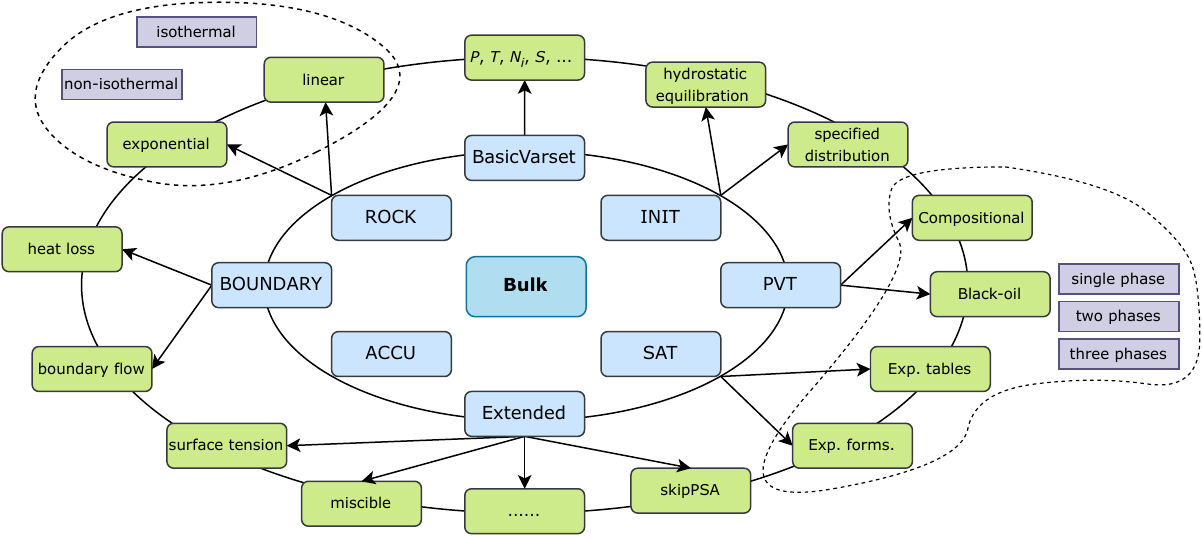}
	\caption{The top-level structure of \texttt{Bulk} module in OpenCAEPoro}
	\label{fig:OCP-Bulk}
\end{figure}

The \texttt{Bulk} module focuses on handling physical processes and computations related to a single grid cell. 
It is the most critical and complex module in the entire software. The top-level structure of \texttt{Bulk} module is illustrated in Figure~\ref{fig:OCP-Bulk}. 
The \texttt{Bulk} module comprises eight components, which are listed as follows:
\begin{enumerate}
	\item \textbf{\texttt{BasicVarset}}: Stores the essential variables necessary for performing basic simulations. These variables, distributed across grid cells, encompass both geometric details of the cells and physical properties of fluid and solid substances. Additionally, the dataset also includes some corresponding derivative variables. The variables in \texttt{BasicVarset} are frequently accessed during the simulation process, with different patterns. Therefore, it is essential to employ a flexible and efficient data structure, specifically the previously introduced SOA format. 
	\item \textbf{\texttt{INIT} module}: Computes the initial state of grid cells, such as pressure, temperature, quantities of components and phases. It supports hydrostatic equilibrium initialization or non-equhydrostaticilibrium initialization with specified initial states for each grid cell.
	\item \textbf{\texttt{PVT} module}: Computes the thermodynamic properties of fluid mixtures within grid cells when reaching equilibrium state. There exist two types of models: compositional models (phase equilibrium calculations based on equations of state) and black oil models (explicit calculations based on experimental tables), which can cover a range of flow scenarios, from single-phase flow to multiphase flow.
	\item \textbf{\texttt{SAT} module}: Computes the permeability properties of phases in fluid mixtures within grid cells, such as relative permeability and capillary pressure. Relevant models are typically based on experimental data and formulations, tailored to accommodate diverse flow scenarios.
	\item \textbf{\texttt{ROCK} module}: Computes the rock properties, such as porosity and absolute permeability, in each grid cell. In cases of non-isothermal cases, the thermal expansion characteristics of rocks must be also taken into account.
	\item \textbf{\texttt{BOUNDARY} module}: Handles physical processes occurring at the reservoir boundaries, such as mass exchange and heat transfer between the reservoir and its external surroundings. These processes typically involve the outermost grid cells of the reservoir. Commonly used conditions include inflow and outflow at constant pressure or constant velocity, as well as heat losses.
	\item \textbf{\texttt{ACCU} module}: Assembles the coefficient matrix for various solution methods, focusing solely on terms related to the grid cells themselves, which means these contributions are positioned at the diagonal elements of the matrix. Typically, \texttt{ACCU} module needs to collaborates with other modules within the \texttt{Bulk} module.
	\item \textbf{\texttt{Extended} module}: Facilitates the extension of program functionality within established simulation frameworks, which is crucial for the extensibility of the program.
\end{enumerate}

\subsubsection{\texttt{Reservoir::BulkConn} and \texttt{Reservoir::AllWells} modules}\label{subsubsec:BulkConn and AllWells}
Within the \texttt{Reservoir} module, the \texttt{BulkConn} module plays a critical role in managing the connectivity between grid cells. 
It is tasked with calculating the mass (or heat) flux across cell connections, driven by processes such as convection, diffusion, or thermal conduction.
Alongside, the \texttt{AllWells} module is specialized in handling the connections between grid cells and wells. 
This includes computing the flux across grid cell-well perforation connections, as well as monitoring the pressure within the well. 
Both modules adhere to the same design philosophy as the \texttt{Bulk} module, endowing the program with enhanced functional extensibility and readability.

\subsubsection{Relationship with solution methods}\label{subsubsec:Relationship with solution methods}
In the field of reservoir numerical simulation, a diverse array of solution methodologies has been developed, each characterized by unique features suited to different problem scenarios. 
We have an ambition to amalgamate different solution approaches within a unified framework, allowing for flexible applications and thereby enabling dynamic collaboration among various methods. 
These features not only have the potential to promote more efficient and robust simulations but also facilitate the development and testing of new methods.
To achieve the goal of effective integration, careful consideration must be given to the programming aspects, particularly the relationship between modules that encapsulates various physical objects and processes (\texttt{Reservoir} module and its numerous sub-modules) and those representing solution methodologies (such as the \texttt{FIM} module and the \texttt{IMPEC} module).	

\begin{figure}[htbp]
	\centering
	\includegraphics[width=0.9\linewidth]{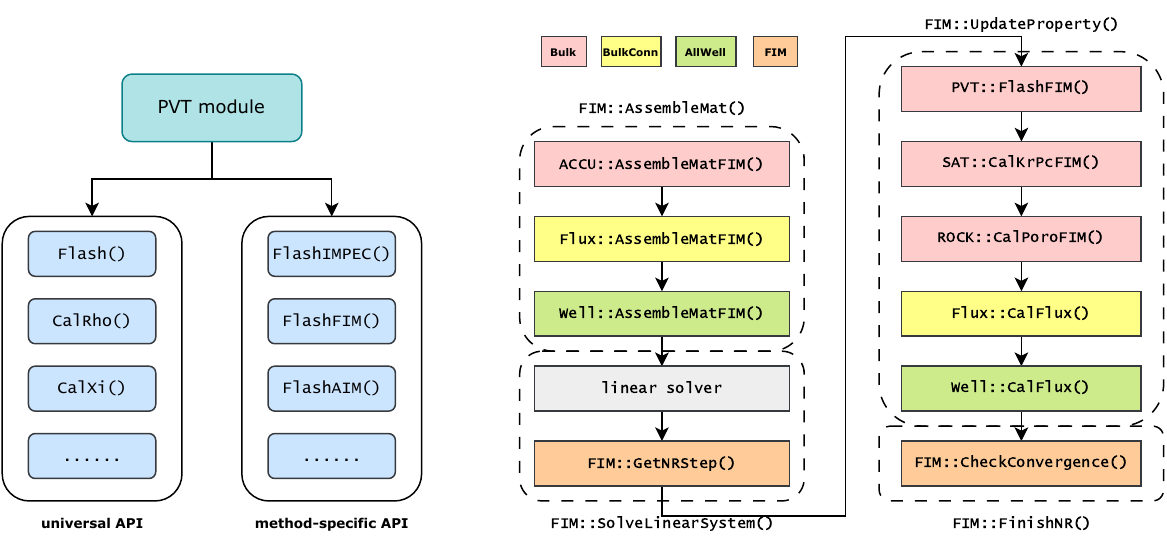}
	\caption{The left diagram illustrates the functional division and related interfaces of the \texttt{PVT} module, while the right diagram schematically demonstrates the application of these method-specific interfaces in the solution process of FIM, where the color of functions within rectangles corresponds to their belonging module.}
	\label{fig:OCP-module-solution}
\end{figure}

For the same simulation problem, all solution methods need to calculate the same reservoir properties. 
The difference lies in the auxiliary variables, such as partial derivatives, used when assembling the Jacobian matrix.
Therefore, for modules like \texttt{Bulk::PVT} and \texttt{Bulk::SAT}, the calculations involved in each method share far more commonalities than differences.
On the other hand, it is essential to maintain the encapsulation of modules, especially those with complex internal architecture.
Preserving their autonomy is crucial for improving the program's maintainability, extensibility and reusability.
Consequently, we opt to confine the computation of physical properties related to this module within itself. Meanwhile, for the next-level modules within \texttt{Reservoir::Bulk}, \texttt{Reservoir::BulkConn}, and \texttt{Reservoir::AllWells}, we divide their functionalities into universal and method-specific categories and provide appropriate interfaces, as shown on the left part of Figure~\ref{fig:OCP-module-solution}.
The universal category includes standard phase equilibrium calculations, phase densities computation, etc. 
While the method-specific group is customized for specific solution methods, such as phase equilibrium calculations for FIM, which require additional computations of related derivative variables on top of the universal ones.
It is foreseeable that \texttt{Flash()} and \texttt{FlashFIM()} share a substantial amount of identical computation processes. 
By integrating these calculations into the \texttt{PVT} module and through thoughtful modularity, redundancy in the code can be minimized effectively. 

The diagram on the right part of Figure~\ref{fig:OCP-module-solution} showcases the application of module-specific interfaces in solution methods.
Using FIM as an example, the flowchart represents the solving process of a nonlinear iteration step, which is divided into four parts: (1) assembling of the Jacobian matrix, (2) solving the linear system and updating the primary variables, (3) updating reservoir properties, and (4) checking the convergence of the nonlinear solution. Within these segments, the solution method is required to organize the FIM-specific interfaces provided by various modules in the \texttt{Reservoir} module to implement the desired solving process.
It is important to note that some details have been simplified or omitted for the sake of clarity and conciseness. 

\subsubsection{\texttt{Extended} modules}\label{subsubsec:Extended}
\begin{figure}[H]
	\centering
	\includegraphics[width=0.9\linewidth]{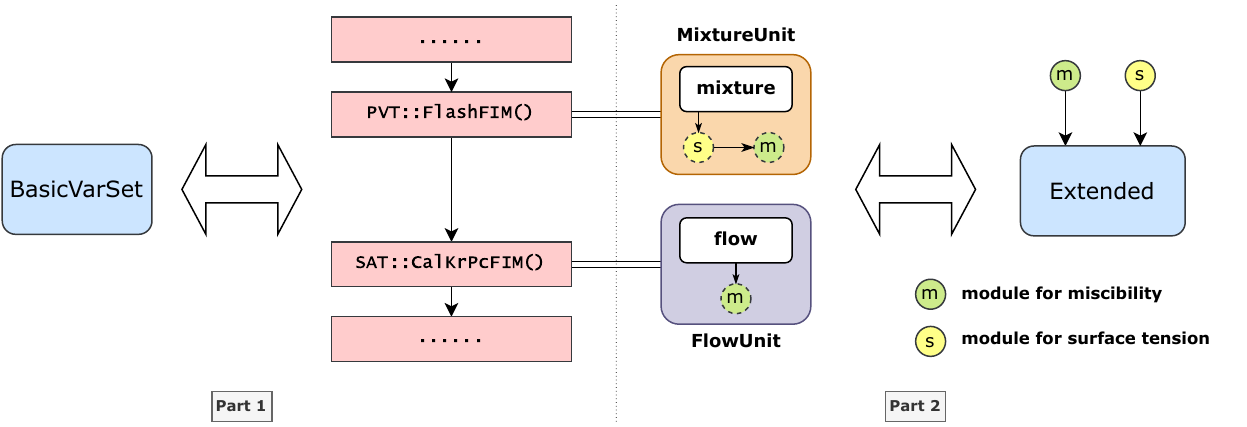}
	\caption{Part 1 illustrates a portion of the main computational process, see Figure~\ref{fig:OCP-module-solution}, and the interaction of these modules with the \texttt{BasicVarSet}. Part 2 demonstrates the localized extension of miscibility functionality through the \texttt{Extended} module.}
	\label{fig:OCP-bulk-work01}
\end{figure}

The \texttt{Extended} modules play crucial roles in the program's functional expansion.
Currently, \texttt{Extended} modules are incorporated within the \texttt{Bulk} module and the \texttt{BulkConn} module.
We use \texttt{Bulk::Extended} module as an example to illustrate the design principles and operation of the \texttt{Extended} module.
Within the \texttt{Bulk} module, we have delineated several principal sub-modules, such as the \texttt{PVT} and \texttt{ACCU} modules, based on physical processes and functions, which can cover most simulation requirements.
Meanwhile, we have included essential variables necessary for a basic simulation, such as pressure $P$, phase saturation $S_{j}$ and phase density $\rho_{j}$, into the \texttt{BasicVarset}. 
Then, these modules extract the required variables for a single grid cell from the \texttt{BasicVarSet}, perform the necessary computations within their internal units, and then return the results to the \texttt{BasicVarSet}, as shown in Part 1 of Figure~\ref{fig:OCP-bulk-work01}.
When integrating new functional modules that involve storing information for all grid cells, we prefer not to alter Part 1, as this would cause the \texttt{BasicVarset} to continuously expand and necessitate constant modifications to the related computational processes and function interfaces. 

Instead, we aim to limit code modifications to the areas where these modules are utilized.
Parts 2 of Figure~\ref{fig:OCP-bulk-work01} illustrate an example of extending the miscibility function in multiphase flow.
One method of calculating the miscibility effect involves using miscibility factors (derived from surface tension between phases) to adjust phase permeability properties, as previously mentioned.
We place the entities of the module for surface tension calculations (denoted as \texttt{s}) and the module for miscibility calculations (denoted as \texttt{m}) in the \texttt{Extended} module, and then assign their pointers to the \texttt{MixtureUnit} module and the \texttt{FlowUnit} module, which are responsible for coordinating calculations of them. 
\texttt{MixtureUnit} module and the \texttt{FlowUnit} module also need to configure these modules during their setup stage.
When a grid cell enters the \texttt{PVT} module, it first completes the calculation of thermodynamic properties through the \texttt{mixture} module and then advances to the \texttt{s} module to calculate surface tension. The \texttt{m} module then determines the miscibility factors with the help of \texttt{s} module and store them within its own variable set. Upon the cell's entering the \texttt{SAT} module, the \texttt{flow} module initially calculates the relative permeability and capillary pressures. Subsequently, the \texttt{m} module modify these properties based on the miscibility factors previously calculated. 
This localized modification allows new modules to be attached to the predefined modules like plugins, enhancing the maintainability of the program.

\subsection{\texttt{Solver} module}\label{subsec:solver}
The \texttt{Reservoir} module encapsulates all information related to the reservoir, including a variety of physical variables and models. 
In an effort to facilitate the integration of diverse solution methods into a unified framework, the functionalities and associated function interfaces of next-level modules within \texttt{Reservoir::Bulk}, \texttt{Reservoir::BulkConn}, and \texttt{Reservoir::AllWells} have been delineated into categories that are either universal or specific to solution methods. 
This strategy effectively transforms the \texttt{Reservoir} module into an extensive library of components for solution methods, where the principal task of these solution methods is to efficiently organize and utilize these components to achieve the intended solving process under a singular framework.

\subsubsection{Hierarchical structure}\label{subsubsec:Hierarchical structure}
\begin{figure}[htbp]
	\centering
	\includegraphics[width=0.9\linewidth]{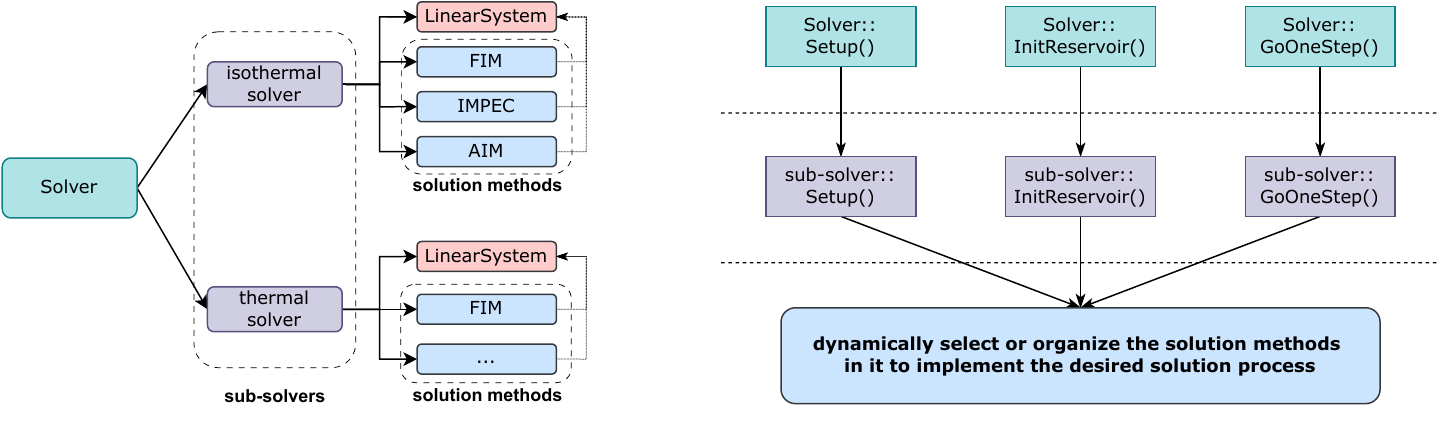}
	\caption{The left figure illustrates the structure of the \texttt{Solver} module, while the right figure depicts the relationships among the \texttt{Solver} module, its sub-solvers, and the methods within these solvers.}
	\label{fig:OCP-Solver}
\end{figure}

As illustrated in Figure~\ref{fig:OCP-Solver}, the \texttt{Solver} module comprises sub-solvers tailored to different problems, such as isothermal and non-isothermal solvers. 
These sub-solvers encompass various solution methods, for instance, IMPEC, FIM, and AIM. 
The \texttt{Solver} module provides a unified interface at the outer layer, with the actual computations being carried out by the sub-solvers. 
Meanwhile, the sub-solvers dynamically select or organize the solution methods beneath them to achieve specific computational objectives.
Additionally, each sub-solver is complemented by a linear system module, which will be discussed in detail later.

The \texttt{Setup} function is designed to prepare the solution methods, encompassing the memory allocation for variables associated with the utilized methods, as well as the configuration and memory allocation for linear solvers. 
If multiple solution methods will be employed in a single simulation, all these methods should be setup at this stage. 
The \texttt{InitReservoir} function is designed for initializing the state of the reservoir. 
The \texttt{GoOneStep} function is dedicated to the resolution of a single time step, constituting the core component within the entire solving process.

\subsubsection{Time-stepping}\label{subsubsec:Time-stepping}
\begin{figure}[htbp]
	\centering
	\includegraphics[width=0.8\linewidth]{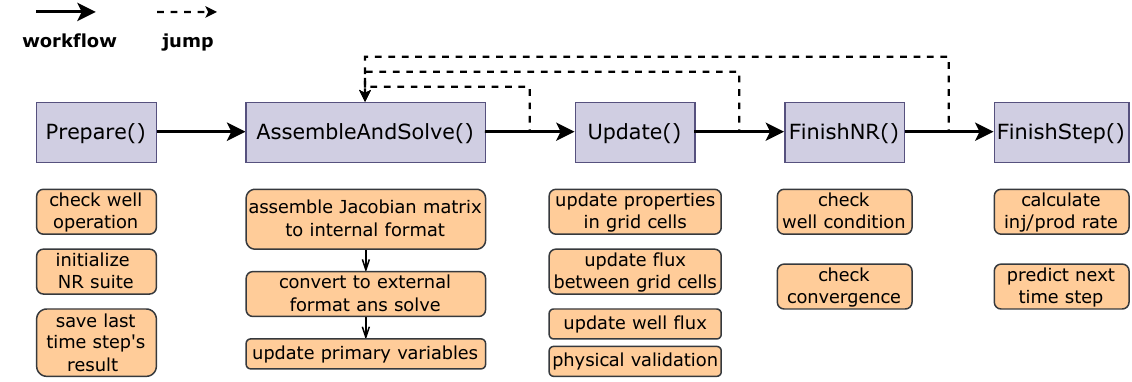}
		\caption{The diagram illustrates the solution flow of a single time step. Each stage is implemented using specific solution methods. Solid arrows guide the direction of the workflow, while dashed arrows represent conditional returns to earlier stages. The yellow boxes indicate the computations that may be included in the corresponding stage.}
	\label{fig:OCP-GoOneStep}
\end{figure}

Regarding the solution of a single time step, the sub-solvers provide a uniform framework for the solving techniques beneath it, as shown in Figure~\ref{fig:OCP-GoOneStep}. 
Indeed, the framework exhibits considerable adaptability, stemming from two fundamental design principles:
\begin{enumerate}
    \item  Solution methods possess a high degree of flexibility in their internal computations and can provide feedback to sub-solvers as needed for transitions between solving stages.
    This capability offers strong adaptability to various methods.
    \item In each stage, sub-solvers dynamically select and organize their solution methods through \texttt{switch-case} statements, thus establishing a foundation for the dynamic switching between methods.
This capability facilitate dynamic collaboration among solution methods.
\end{enumerate}

\begin{figure}[H] 
    \centering
    \begin{subfigure}[b]{\linewidth}
        \centering
        \includegraphics[width=0.9\linewidth]{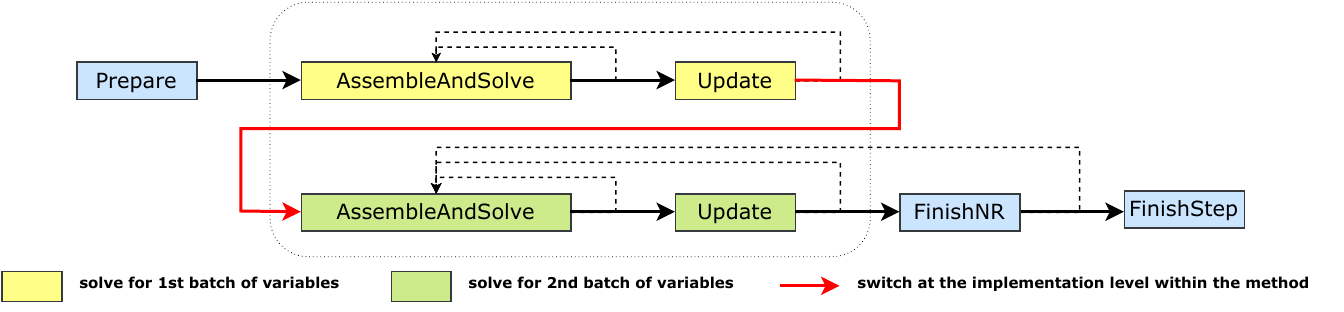}
        \caption{}
        \label{fig:extension:SIM}
    \end{subfigure}

    \begin{subfigure}[b]{\linewidth}
        \centering
    \includegraphics[width=0.9\linewidth]{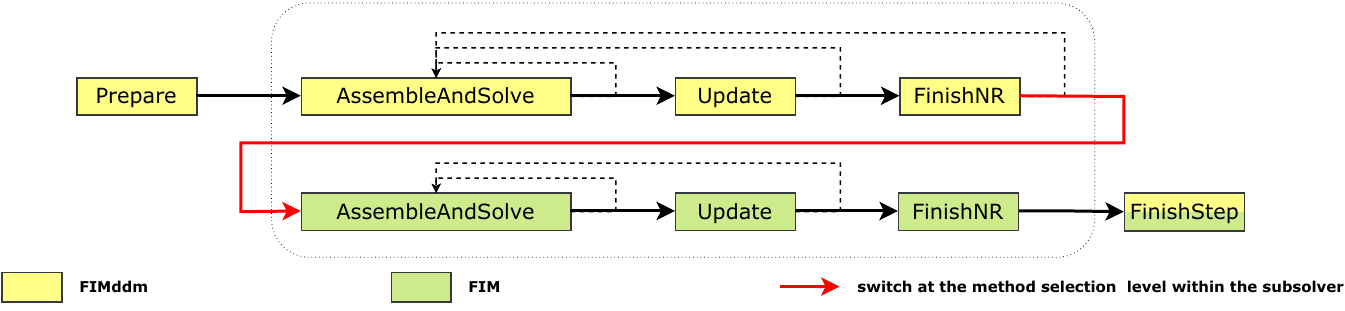}
        \caption{}
        \label{fig:extension:FIMddm}
    \end{subfigure}
 
    \caption{Two extensions within the GoOneStep framework are as follows: (a) for the Sequential Fully Implicit (SFI) method, and (b) for method collaboration, where FIMddm is used to provide the initial guess for FIM.}
    \label{fig:OCP-GoOneStep:extension}
\end{figure}

We now showcase the adaptability of the framework depicted through a couple of examples, as illustrated in Figure~\ref{fig:OCP-GoOneStep:extension}.
The first example demonstrates an adaptation of the Sequential Fully Implicit (SFI) method~\parencite{jenny_adaptive_2006}, showcasing the flexibility of computational content at each stage and method-driven stage transitions. 
The SFI method is characterized by its batch-wise, implicit processing of primary variables, which necessitates the sequential resolution of multiple nonlinear systems within a single Newton iteration, consequently requiring the solution of multiple linear systems.
In this case, we assume that the number of batches of primary variables is two.
After finishing the computations for the 1st batch of variables, the program returns to the previous stage to process the 2nd batch of variables, thereby concluding a full Newton iteration.
The second example highlights the collaboration between solution methods, particularly in using one method to provide an initial guess for another. 
This extension showcases the coordinating role of sub-solvers in different solution methods and the sub-solver-driven stage transitions.
Here, we employ a local method (FIM based on the domain decomposition method, abbreviated as FIMddm, which will be detailed in the following section) with lower convergence performance but better parallel scalability to find a suitable initial solution for a global method (FIM). 
Once FIMddm meets the specified convergence criteria, the sub-solver switches the solution method to FIM.
It should be noted that the actual computational framework remains as described in Figure~\ref{fig:OCP-GoOneStep}; this representation is merely to better illustrate the transitions between solution stages.

\subsubsection{\texttt{LinearSystem} module}\label{subsubsec:LinearSystem}
In the previous subsection, we have shown the adaptability of the solution framework within \texttt{GoOneStep} by two examples. 
A notable aspect shared by these examples is employing sparse matrices of diverse types in each time step, including variations in block sizes (as observed in the first example, the number of degrees of freedom within grid cells may vary between the two batches) and discrepancies in dimensions (as noted in the second example, the matrices assembled by each process in FIMddm may be of square or rectangular shape with varying numbers of columns. 
In contrast, in the FIM, the matrices are rectangular with a constant number of rows and columns). 
Furthermore, distinct solver packages tailored to different scenarios may be employed within a single simulation run. Consequently, a flexible and efficient linear system module is crucial.

\begin{figure}[htbp]
	\centering
	\includegraphics[width=0.9\linewidth]{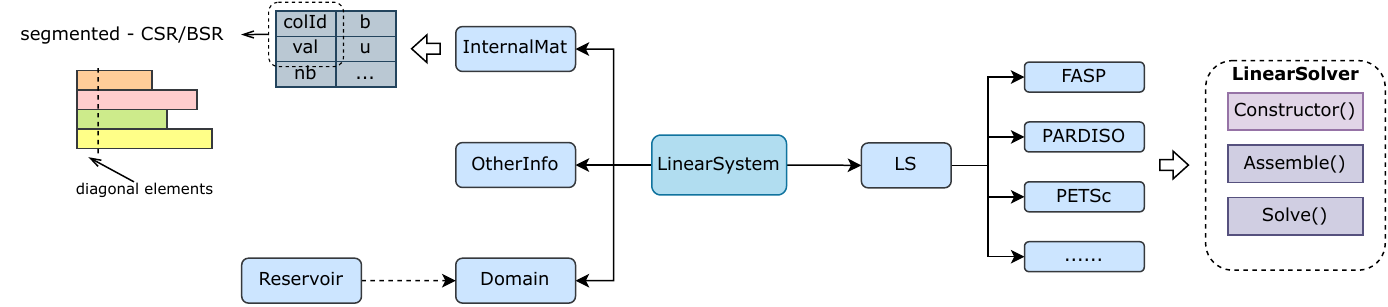}
	\caption{Linear system module in OpenCAEPoro.}
	\label{fig:OCP-LS-Module}
\end{figure}

Figure~\ref{fig:OCP-LS-Module} presents the architectural diagram of the \texttt{LinearSystem} module, comprising four major components:

\begin{enumerate}
	\item \textbf{\texttt{InternalMat}}: Dedicated to encapsulating the essential data of the linear system, such as the Jacobian matrix $A$, right-hand side vector $b$, solution vector $u$, and dimension of sub-blocks $nb$, which may vary during the simulation. In this architecture, $b$ and $u$ are formatted as one-dimensional \texttt{vector}, whereas $A$ is organized in a segmented CSR/BSR format. This strategy involves dividing the original CSR/BSR framework along the matrix rows and integrating these segments into a two-dimensional \texttt{vector}, as illustrated on the left of Figure~\ref{fig:OCP-LS-Module}. As a result, the column indices (\texttt{colId}) and values (\texttt{val}) of the non-zero elements are both structured as two-dimensional \texttt{vector}, with each row corresponding to the respective row in the matrix. This structure is better suited for matrices with dynamically changing sparse patterns and block sizes, and it eliminates the need for the array traditionally used to denote the starting positions of non-zero elements in each row in conventional CSR/BSR formats.
	\item \textbf{\texttt{LS}}: An array of pointers to all the subclass instances of linear solvers in use. The specific element of this array being used is determined by a marker index.
	These subclasses, deriving from the abstract base class \texttt{LinearSolver}, implement a range of solver packages, including FASP~\parencite{fasp_development_team_fasp_2023}, PARDISO~\parencite{schenk_pardiso_2001}, and PETSc~\parencite{balay_petsc_2015}.
	In their constructors, these subclasses are tasked with the memory allocation for vital variables, such as the matrix in its specific format, and setting up solution parameters, such as tolerance and maximum iterations. These parameters can also be imported from external files.
	The base class \texttt{LinearSolver} also furnishes a standardized external interface for these subclasses. 
	The \texttt{Assemble()} function is designed to transform the internal matrix, known as \texttt{InternalMat}, into the requisite format for specific solver packages. 
	The \texttt{Solve()} function triggers the specific solving functions of the related solver package, returning the solution and iteration count.
	\item \textbf{\texttt{Domain}}: Stores the communication information between processes. Herein, The \texttt{Domain} module is employed to aid in computing global information for the linear system, such as the global indices of matrix elements, when the internal matrix is converted to the format required by external solvers. It is stored as a pointer in \texttt{LinearSystem}, while the entity resides in the \texttt{Reservoir} module.
	\item \textbf{\texttt{OtherInfo}}: Additional information, such as working indices identifying the specific instance of the \texttt{LS} array currently in use.
\end{enumerate}
It should be noted that, due to the \texttt{LinearSystem} module being designed at the same level as the solution methods, different solution methods can share it within a simulation, particularly its \texttt{InternalMat}, thereby reducing unnecessary memory overhead.

\subsection{Parallelization}\label{subsec:Parallelization}
\subsubsection{Parallel strategy}\label{subsubsec:Parallel strategy}
\begin{figure}[htbp]
	\centering
	\includegraphics[width=0.9\linewidth]{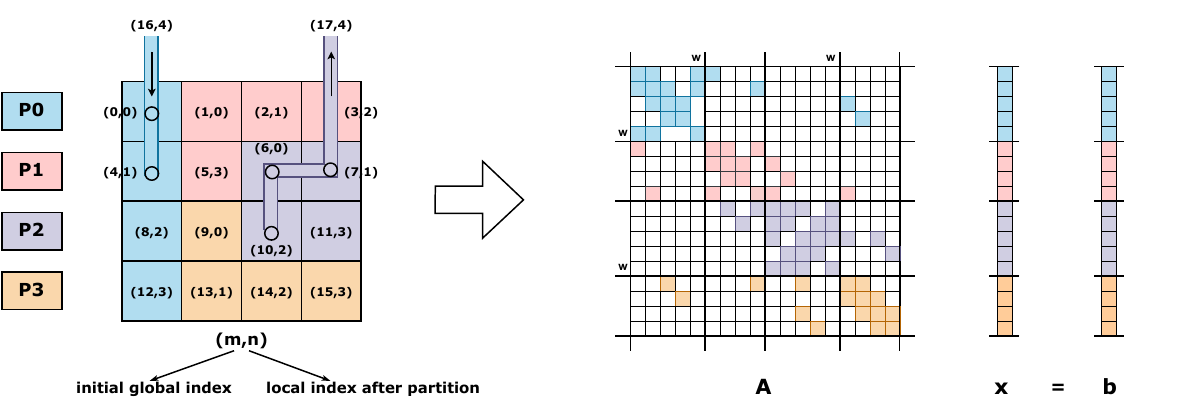}
	\caption{A schematic diagram of the underlying strategy of parallel simulation in OpenCAEPoro. The computational area consists of a $4\times4$ grid and 2 wells, segmented by four processes. Each cell's association with a process is denoted by a color corresponding to that process. In addition, each process's owned portion of the linear system corresponds to its assigned color, as shown on the right. The symbol $w$ denotes the rows and columns where the well cells are located.}
	\label{fig:OCP-parallel01}
\end{figure}

The underlying strategy of parallel multiphase multicomponent flow simulation is to divide the computational domain into multiple sections, with each section managed by one or several processes. 
During the nonlinear solving process, all subdomains collectively form a global Jacobian matrix, which is solved jointly by all processes, as demonstrated in Figure~\ref{fig:OCP-parallel01}.
Ideally, if the computational load and the number of neighbors for each grid cell are uniform, the distribution of grid cells across processes should be nearly equal to maintain load balance.
Additionally, the well-perforated grid cells are required to be in the same process as this well, facilitating the sequential handling of the intricate and diverse well items.

As depicted on the right part of Figure~\ref{fig:OCP-parallel01}, each process is responsible for assembling the row blocks of the Jacobian matrix for the cells (referred to as interior cells of current process) assigned to it. 
Consequently, in order to compute the off-diagonal blocks in the Jacobian matrix, a process requires information of not only the interior cells, but also the ghost cells (those cells that connect to the interior cells but are distributed across other processes). 
This implies the necessity for inter-process communication to retrieve the relevant data of ghost cells. 
However, the intricacies of calculating Jacobian matrix elements depend significantly on the specific problem, the model employed, and the solution method utilized, leading to variations in the type and quantity of variables exchanged between processes for different cases. 
In OpenCAEPoro, each process initially acquires all essential information about the interior cells and ghost cells, subsequently exchanges only primary variables, such as pressure $P$, temperature $T$, and composition $N_{i}$, within the ghost cells after these variables have been updated. 
These variables are then utilized to calculate the required physical information for all ghost cells, which will be employed to assemble the Jacobian matrix.

\subsubsection{Inter-process communication mechanism}\label{subsubsec:Inter-process communication mechanism}
Communications between processes during a simulation primarily originates from two aspects. 
The first involves synchronization of global information, such as convergence checks and physical validation, which can often be accomplished using operations like \texttt{MPI\_Allreduce()}; 
The second aspect is the exchange of information between processes regarding ghost cells, which requires point-to-point communication between processes.

In OpenCAEPoro, we have developed an efficient communication mechanism in the \texttt{Domain} module, to address point-to-point communications between processes.
We begin by reordering the interior cells. As shown on the left side of Figure~\ref{fig:OCP-parallel01}, each cell is marked by a tuple of values $(m,n)$, where $m$ denotes the initial global index of this cell before grid partitioning, and $n$ signifies the cell's local index within its respective process after partitioning, which also serve as the row and column indices for the cells in the local matrix depicted on the right side of Figure~\ref{fig:OCP-parallel01}. 
For two interior cells $(m_{1},n_{1})$ and $(m_{2},n_{2})$, if 
$m_{1}>m_{2}$, then $n_{1}>n_{2}$; that is, within each process, the interior cells are arranged in ascending order according to their initial global indices.
It should be highlighted that well cells are always interior cells and are contained within the \texttt{AllWells} module.

\begin{figure}[htbp]
	\centering
	\includegraphics[width=0.8\linewidth]{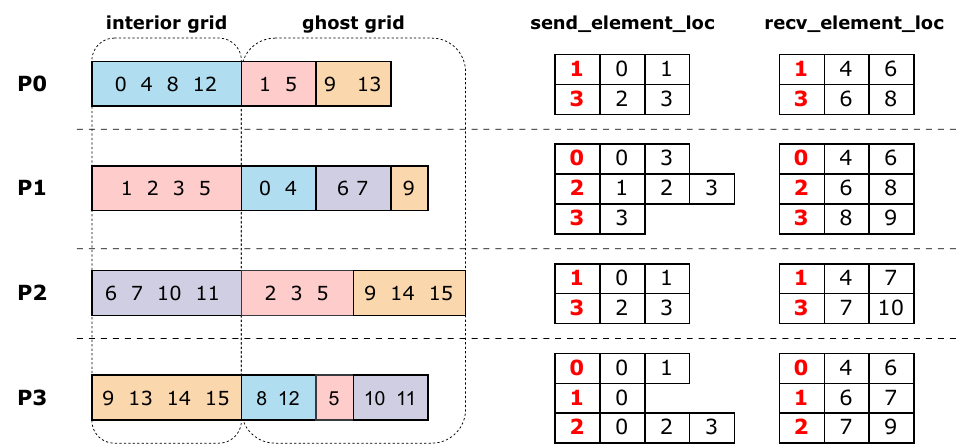}
	\caption{The left figure illustrates the storage order of grid cells within a process, with color blocks mathcing the process colors of these cells, as depicted in Figure~\ref{fig:OCP-parallel01}. For the two \texttt{map} structures in the right figure, the \texttt{key}, marked in bold red, corresponds to the global ranks of neighboring processes, while the subsequent \texttt{value} records the grid locations.}
	\label{fig:OCP-parallel05}
\end{figure}

Subsequently, our focus will shift to the local organization of grid cells stored in the \texttt{Bulk} module, encompassing both interior and ghost grid cells.
As illustrated on the left side of Figure~\ref{fig:OCP-parallel05}, we first arrange the interior grid cells at the forefront, sorted as previously described.
Following this, ghost grid cells are ordered in ascending order based on the rank of their associated process, with ghost cells from the same process further organized in ascending order of their initial global indices.
For instance, within process \texttt{P1}, we first arrange its interior grid cells, which have initial global indices of 1, 2, 3, and 5. Then, we sequence the ghost grid cells from process \texttt{P0}, followed by those from \texttt{P2}, and finally from \texttt{P3}. All these cells are arranged in ascending order of their initial global indices within their respective segments.

Now, we can conveniently represent the information for point-to-point communication between processes using two \texttt{map} structures: \texttt{send\_element\_loc} and \texttt{recv\_element\_loc}, represented as \texttt{map<int,set<int>}\texttt{>} and \texttt{map<int,vector<int>}\texttt{>}, respectively, as illustrated on the right side of Figure~\ref{fig:OCP-parallel05}. 
The \texttt{send\_element\_loc} holds sending-related information for the current process, indicating which grid cells' information is to be sent to which processes. 
Conversely, the \texttt{recv\_element\_loc} contains receiving-related information for the current process, specifying which grid cells' information will be received from which processes.  
In \texttt{send\_element\_loc}, the sending locations are identified by a set of local indices in ascending order, while in \texttt{recv\_element\_loc}, the receiving locations are determined by index ranges, because the order of received grid information always continuously corresponds to the storage order. 
For example, process \texttt{P1} sends information of grid cells with local indices 0 and 3 to process \texttt{P0}. When process \texttt{P0} receives the information, it will place them sequentially in the grid cells within local index range $[4,6)$.

These two \texttt{map} serve as a repository for point-to-point communication between processes. Using them, we can easily implement point-to-point communication, such as updating the primary variables of ghost grid cells to calculate the elements of the Jacobian matrix or computing the global indices of these elements.

\section{Adaptively coupled domain decomposition method}\label{Adaptive Coupled Domain Decomposition Method}
As shown in the previous section, a parallel solving strategy has been employed where all subdomains are solved in a coupled manner. 
Consequently, the nonlinear systems obtained through this method are identical to those arising from serial solving, with no loss of information despite the subdomains being distributed across different processes.
However, it incurs significant computational costs during the linear solving phase due to the solution of the global linear system. 
This problem becomes more serious when solving with large number of processes (where tens of thousands of cores are employed).
A potentially effective approach involves the use of domain decomposition methods (DDM), where each process independently solves its assigned sub-domain~\parencite{toselli2004domain}. 
This method can achieve exceptionally high parallel efficiency; however, its convergence performance often struggle due to neglecting important coupling relationships between subdomains.  
As a result, this method can be strategically employed to provide initial solutions for the globally coupled method during the nonlinear solving iterations. 
This can reduce the number of global Newton iterations and, subsequently, the global linear iterations, thereby accelerating the simulation process. 

In OpenCAEPoro, we have developed a general method based on the domain decomposition concept, referred to as adaptively coupled domain decomposition methods. 
In contrast to the classical methods, adaptively coupled domain decomposition methods improve convergence performance by adaptively identifying significant couplings between subdomains and solving these subdomains in a locally coupled manner during the simulation.
On the other hand, these coupled regions are often fragmented and occupy a very small area compared to the total simulation domain. Consequently, the number of processes involved in coupled computations is minimal, maintaining high parallel efficiency. This provides a significant advantage for large-scale parallel computations, especially when the global linear solver reaches its parallel bottleneck.

These significant couplings typically exhibit noticeable characteristics. 
For example, in reservoir problems, they are often located at the moving front of phase interfaces, where changes in phase saturation are greater compared to other areas. 
Various methods can be employed to establish adaptive coupling relationships between subdomains. One such method, which will be used in Section 5, is as follows:
\begin{enumerate}
    \item At the beginning of each time step, mark the grid cells with significant phase saturation changes from the previous time step, such as those with $\Delta S >5\times10^{-3}$.
    \item Establish a coupling relationship graph between subdomains:
    \begin{enumerate}
        \item If the current subdomain contains marked interior grid cells, then it is coupled with all its neighboring subdomains.
        \item If the current subdomain contains marked ghost grid cells, it is coupled with the neighboring subdomains where these ghost grid cells are distributed.
    \end{enumerate}
    \item Use Boost Graph Library~(BGL) to find all connected components of the coupling graph~\parencite{siek_boost_2002}.
\end{enumerate}

\begin{figure}[htbp]
	\centering
	\includegraphics[width=0.8\linewidth]{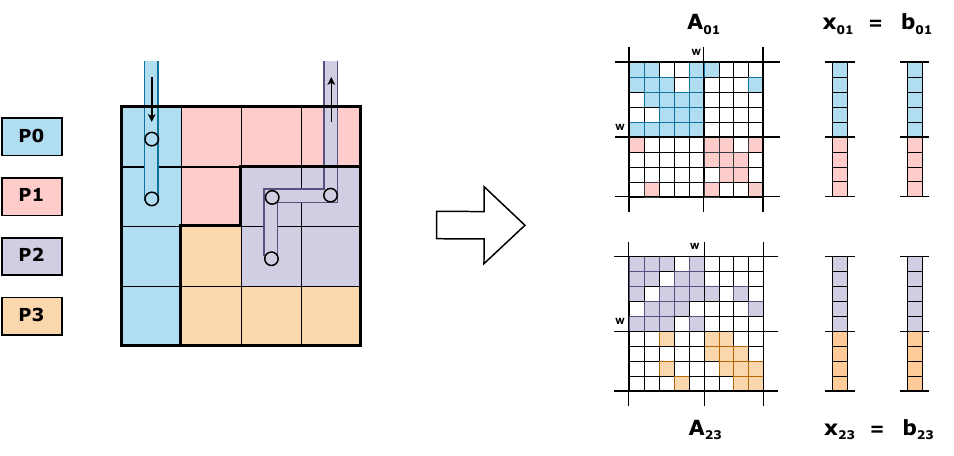}
	\caption{Schematic of a generalized DDM-based method, where Process \texttt{P0} and \texttt{P1}, as well as Process \texttt{P2} and \texttt{P3}, are separately coupled for solving. 
    }
	\label{fig:OCP-parallel07}
\end{figure}

Figure~\ref{fig:OCP-parallel07} illustrates the schematic of the solution after the coupling pattern has been determined, where subdomains managed by processes \texttt{P0} and \texttt{P1} are coupled, as well as those managed by processes \texttt{P2} and \texttt{P3}. 
Consequently, processes \texttt{P0} and \texttt{P1} collaboratively solve the nonlinear system generated in the coupled region they are responsible for, thus solving the linear system $A_{01}x_{01}=b_{01}$ jointly, similar to processes \texttt{P2} and \texttt{P3}.
The boundary conditions between these coupled regions can be strategically selected as either constant pressure, constant flow rate, or other more complex conditions.
Takeing FIM as an example, if all subdomains are solved in a coupled manner, the method is global FIM. Conversely, if every subdomain is solved independently, it results in the classical DDM-based FIM. 

\begin{figure}[htbp]
	\centering
	\includegraphics[width=0.8\linewidth]{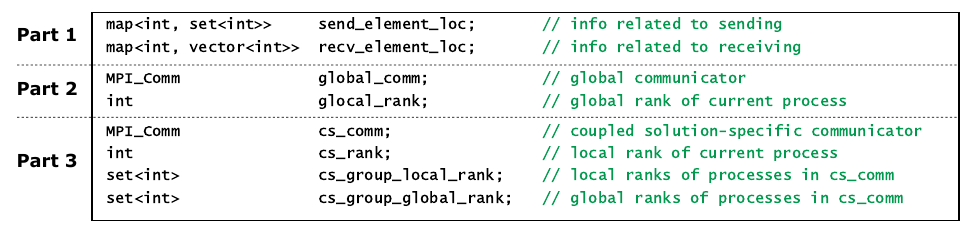}
	\caption{Three types of communication information in the \texttt{Domain} module}
	\label{fig:OCP-parallel08}
\end{figure}

Such extensions can be easily implemented in OpenCAEPoro. 
Within the \texttt{Domain} module, communication-related information is divided into three parts, as shown in Figure~\ref{fig:OCP-parallel08}. 
\texttt{Part~1}, as previously described, includes two core data structures that facilitate point-to-point inter-process communication. 
\texttt{Part~2} stores information related to global communication, while \texttt{Part~3} focuses on communication for dynamic coupled solving. Importantly, \texttt{Part~3} is tasked with providing communication information to the linear system. It should be noted that in scenarios of globally coupled solving, the information in \texttt{Part~3} becomes global.

At the beginning of each time step, processes group together adaptively (for example, using the previously mentioned method), setting up the information for \texttt{Part~3}. Through \texttt{Part~1} and \texttt{Part~3}, the processes can determine the local-global indices (global indices relevant within the locally coupled group of subdomains) of non-zero matrix elements. Additionally, within the ghost grid cells, the current process can identify which parts act as degrees of freedom in the locally coupled system and which parts serve as boundaries between coupled subdomains for imposing boundary conditions.
Armed with these information, processes can be locally coupled to solve their respective nonlinear systems.
When all locally coupled nonlinear system satisfy the convergence criteria, the solution process concludes. 
This method can be employed to provide initial solutions for the global method, thereby accelerating its convergence.

\section{Numerical Experiments}\label{sec:Numerical Experiments}
\subsection{Experimental environment}\label{subsec:Experimental environment}
Our testing platform comprises personal laptops and cluster of the Beijing Super Cloud Computing Center. Validation experiments, being relatively small in scale, were run sequentially on personal laptops. In contrast, tests related to parallel scalability and adaptively coupled DDM were conducted on the Beijing Super Cloud Computing Center. The former tests were performed in the BSCC-T cluster region, where each node has 96 cores, while the latter were carried out in the BSCC-JN cluster, where each node has 56 cores.

We utilized CPR algorithms as the linear solver. 
For validation tests, we employed FASP~\parencite{fasp_development_team_fasp_2023}. 
In the latter two tests, we used PETSc~\parencite{balay_petsc_2015} and Hypre\parencite{hypre}, where the AMG algorithm was implemented using Hypre's Boomer-AMG, and the overall LU smoothing was achieved with Block-Jacobian ILU(0). The iterative method used was FGMRES.

\subsection{Validation tests}\label{subsec:Validation tests}
We validate the reliability of the program through seven SPE comparative solution projects and a simple example with a corner-point grid. This includes the two-phase black oil models: SPE6~\parencite{firoozabadi_sixth_1990}, SPE10~\parencite{christie_tenth_2001}, and CP; the three-phase black oil models: SPE1~\parencite{odeh_comparison_1981} and SPE9~\parencite{killough_ninth_1995}; the compositional models: SPE3~\parencite{kenyon_third_1987} and SPE5~\parencite{killough_fifth_1987}; and the thermal recovery model: SPE4~\parencite{aziz_fourth_1987}. Additionally, the SPE6 case involves a dual-porosity medium, and the CP case example uses a corner-point grid.
The employed solution methods involve IMPEC, FIM, and AIM, which yielded consistent results. By comparing our results with a commercial software, we further verified the program's reliability.
For the scenario of thermal recovery and dual-porosity media, we have currently only implemented the FIM. Consequently, we present only the FIM results for the SPE4 and SPE6 cases. Additionally, for the SPE10 case, due to the complexity arising from the highly heterogeneous nature of the reservoir medium, we opted to use only the FIM for testing.
\begin{figure}[H] 
    \centering

    \begin{subfigure}{0.26\textwidth}
        \centering
        \includegraphics[width=\linewidth]{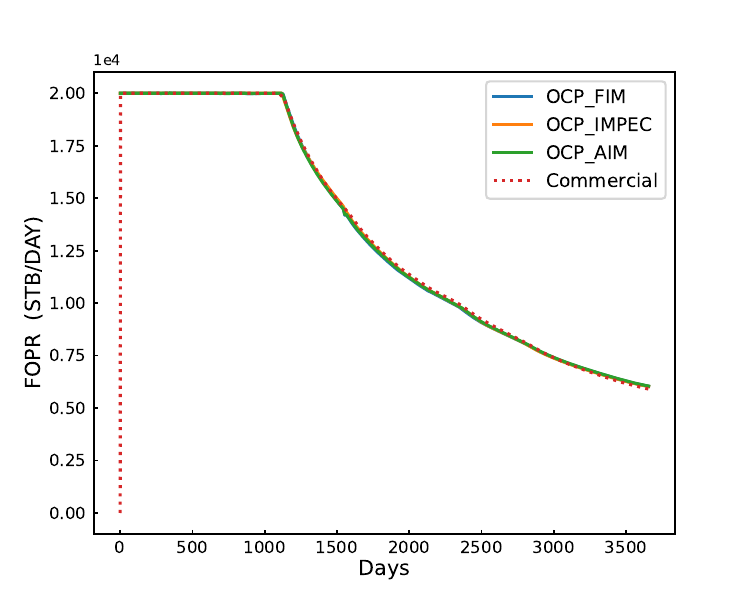}
        \caption{SPE1}
    \end{subfigure}
    \hspace{-0.03\textwidth}
    \begin{subfigure}{0.26\textwidth}
        \centering
        \includegraphics[width=\linewidth]{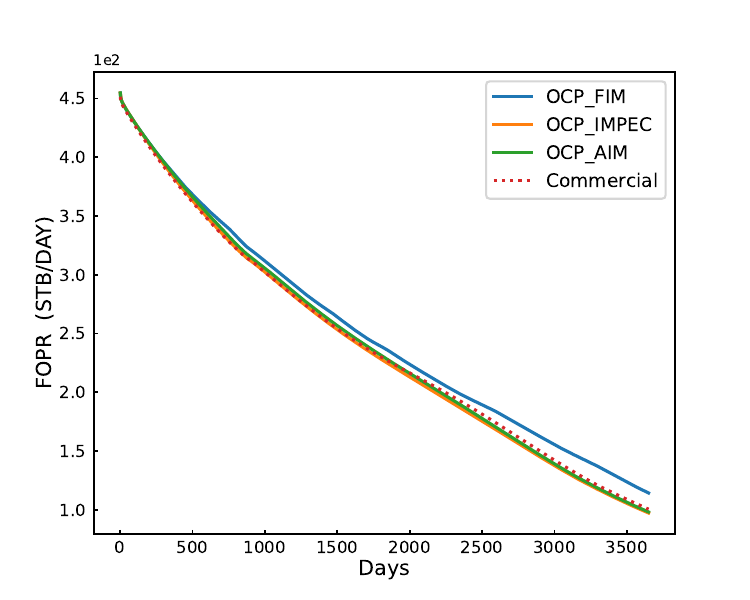}
        \caption{SPE3}
    \end{subfigure}
    \hspace{-0.03\textwidth}
    \begin{subfigure}{0.26\textwidth}
        \centering
        \includegraphics[width=\linewidth]{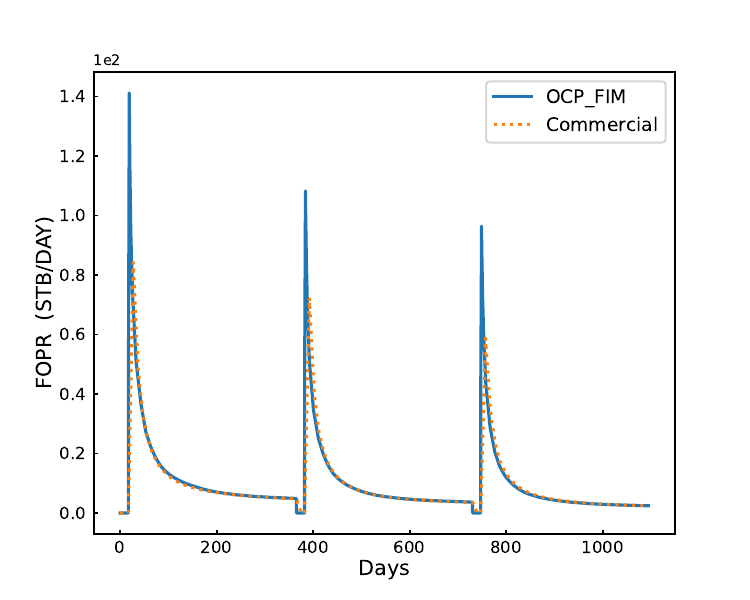}
        \caption{SPE4}
    \end{subfigure}
    \hspace{-0.03\textwidth}
    \begin{subfigure}{0.26\textwidth}
        \centering
        \includegraphics[width=\linewidth]{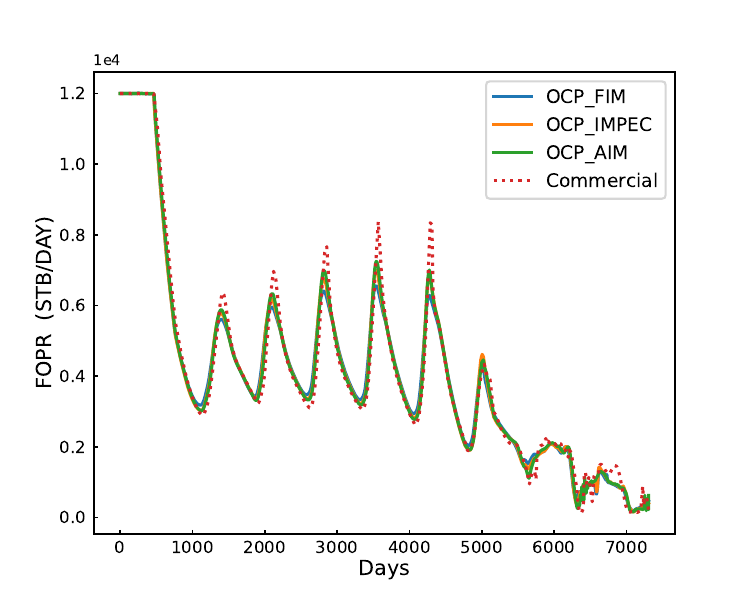}
        \caption{SPE5}
    \end{subfigure}

    \begin{subfigure}{0.26\textwidth}
        \centering
        \includegraphics[width=\linewidth]{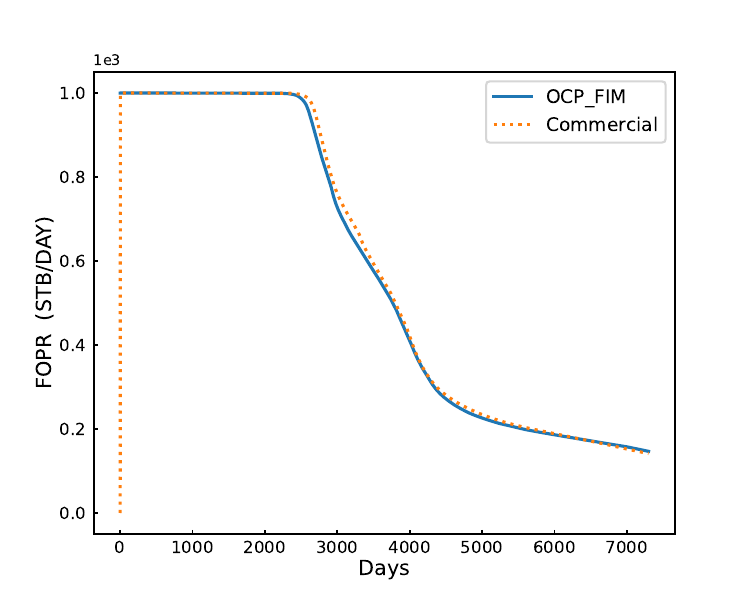}
        \caption{SPE6}
    \end{subfigure}
    \hspace{-0.03\textwidth}
    \begin{subfigure}{0.26\textwidth}
        \centering
        \includegraphics[width=\linewidth]{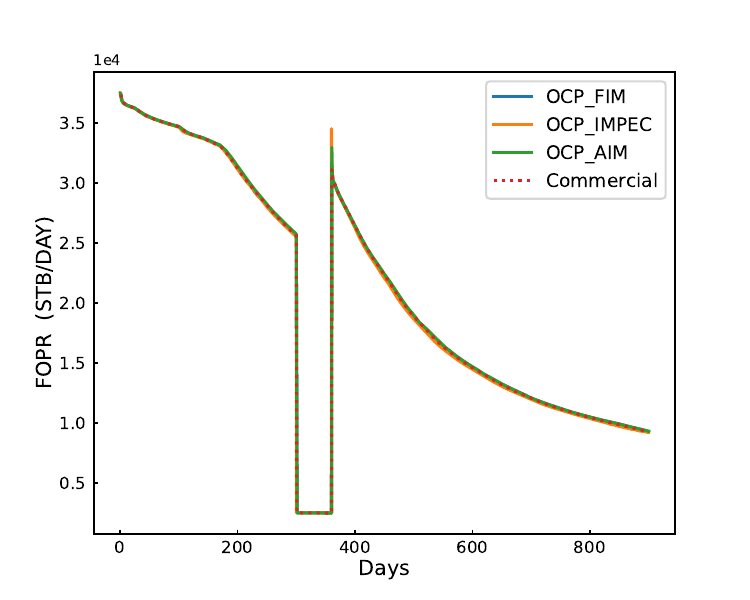}
        \caption{SPE9}
    \end{subfigure}
    \hspace{-0.03\textwidth}
    \begin{subfigure}{0.26\textwidth}
        \centering
        \includegraphics[width=\linewidth]{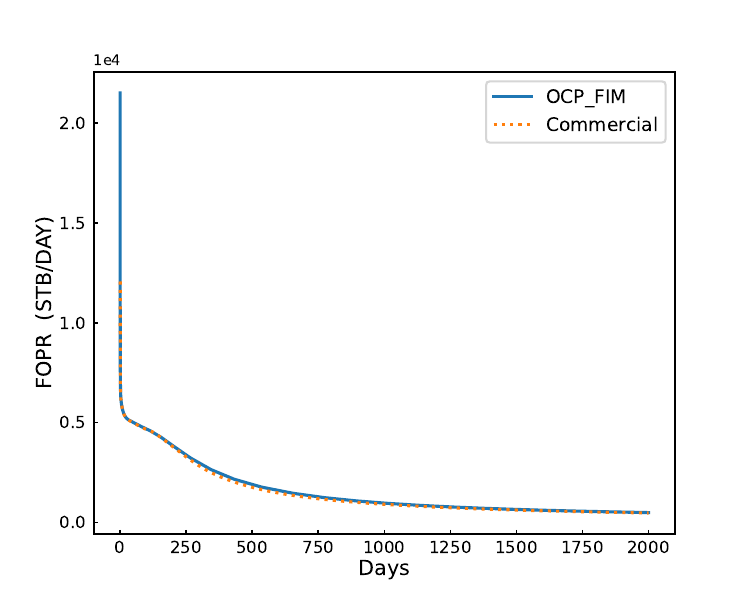}
        \caption{SPE10}
    \end{subfigure}
    \hspace{-0.03\textwidth}
    \begin{subfigure}{0.26\textwidth}
        \centering
        \includegraphics[width=\linewidth]{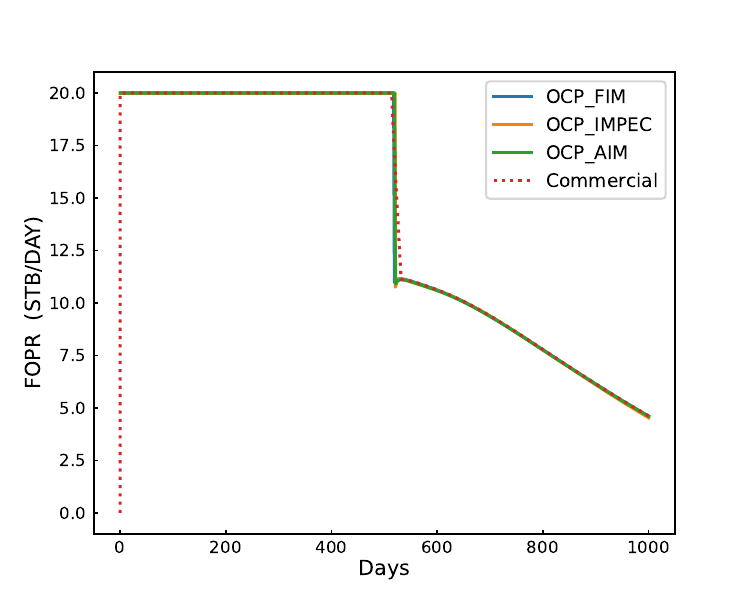}
         \caption{CP}
    \end{subfigure}
 
    \caption{Solutions computed by OpenCAEPoro and commercial software for seven SPE benchmark cases and one simple case with a corner-point grid.}
    \label{fig:validation tests}
\end{figure}

\subsection{Parallel scalability tests}\label{subsec:Parallel scalability tests}
For our parallel scalability testing, we have modified and extended the SPE5 benchmark case.
Our primary objective is to assess the computational performance of the program in parallel scenarios. 
It is crucial to maintain a consistent number of nonlinear and linear iterations across various test cases to ensure the comparability of results. 
This is achieved by ensuring that the problems retain the same characteristics and level of difficulty, irrespective of their scale.
While increasing the grid resolution can indeed scale up the problem size, it also inherently raises the problem's complexity, thereby necessitating smaller time steps. 
To avoid this complication, we opted to increase the number of grids without altering their size. This approach allows us to scale the problem size while preserving its original complexity, facilitating a more accurate evaluation of parallel computational performance.

Specifically, we constructed an initial grid based on the SPE5 benchmark case and then increased the number of grids in the horizontal directions ($x$ and $y$ directions) while maintaining a constant grid count in the vertical direction ($z$ direction).
The final grid size is $781\times781\times6$, which equates to approximately 3.7 million cells and a total of 29 million degrees of freedom (8 degrees of freedom per cell, consisting of 7 component mole number variables and 1 pressure variable). 
Simultaneously, we modified the well configuration to the classic five-spot pattern, consisting of one injection well and four production wells.
In terms of the solution method, we employed Fully Implicit Method (FIM) with a convergence residual of $10^{-3}$. 
The convergence residual for linear solver is set to $10^{-3}$. 
The total simulation time is 100 days.

Table~\ref{tab:strong_parallel} presents the performance results, listing the most time-consuming parts.
\texttt{np} denotes the number of employed processes. \texttt{NRiters} and \texttt{LSiters} represent the total number of iterations for Newton-Raphson method and linear solver, respectively. 
\texttt{Update} refers to the time spent in the update of reservoir properties (see Figure~\ref{fig:OCP-GoOneStep}). \texttt{Assembling} represents the time required for assembling the Jacobian systems. 
\texttt{Linear Solver} denotes the time taken for solving the linear system, and \texttt{Total} represents the overall wall time. 
It should be noted that the \texttt{Update} stage is almost entirely local, except for requiring a global reduction operation due to the synchronization of the results of physical checks. 
The \texttt{Assembling} part is completely local. Consequently, both stages exhibit good parallel strong scalability.

\begin{table}
  \centering
  \caption{Parallel strong scalability results for the extended SPE5 benchmark case. The numbers in parentheses indicate the speedup ratio relative to the first row.}
  \label{tab:strong_parallel}
  \scalebox{0.9}{
  \begin{tabular}{ccccccc}
    \toprule
    np & NRiters & LSiters & Assembling (s) & Linear Solver (s) & Update (s) & Total (s)\\
    \midrule
    1 & 47 & 155 & 848\phantom{00.}(1.0) & 8978\phantom{0}(1.0) & 1483\phantom{00.}(1.0) & 11620\phantom{0}(1.0) \\
    2 & 47 & 154 & 433\phantom{00.}(2.0) & 5085\phantom{0}(1.8) & 738\phantom{000.}(2.0) & 6441\phantom{00}(1.8) \\
    4 & 47 & 160 & 232\phantom{00.}(3.7) & 2745\phantom{0}(3.3) & 374\phantom{000.}(4.0) & 3471\phantom{00}(3.4) \\
    8 & 47 & 161 & 111\phantom{00.}(7.6) & 1428\phantom{0}(6.3) & 190\phantom{000.}(7.8) & 1791\phantom{00}(6.5) \\
    16 & 47 & 161 & 66\phantom{00.}(12.9) & 769\phantom{0}(11.7) & 110\phantom{00.}(13.5) & 968\phantom{00}(12.0) \\
    32 & 47 & 161 & 29\phantom{00.}(28.9) & 406\phantom{0}(22.1) & 52\phantom{000.}(28.5) & 510\phantom{00}(22.8) \\
    64 & 47 & 162 & 20\phantom{00.}(42.6) & 250\phantom{0}(35.6) & 30\phantom{000.}(48.7) & 315\phantom{00}(36.7) \\
    128 & 47 & 163 & 8.9\phantom{00}(95.3) & 257\phantom{0}(34.9) & 17\phantom{000.}(86.8) & 296\phantom{00}(39.3) \\
    256 & 47 & 172 & 4.6\phantom{0}(182.3) & 273\phantom{0}(32.3) & 9.2\phantom{00}(161.8) & 295\phantom{00}(39.4) \\
  \bottomrule
\end{tabular}}
\end{table}

\subsection{Adaptively coupled DDM}\label{subsec:Adaptively coupled DDM}
We present a simple example to demonstrate the advantages of the Domain Decomposition Method (DDM) utilizing the strategy of adaptively coupled solutions. 
We selected the first layer of the benchmark case SPE1 as the computational domain, resulting in an essentially two-dimensional simulation that better demonstrate the concept of adaptively coupled method. 
The grid was refined and extended to a final size of $4000\times4000\times1$, with each grid cell measuring $20\text{ft}\times20\text{ft}\times20\text{ft}$. Additionally, we arranged four five-spot well patterns. The positions coordinates of the four injection wells are $(1000, 1000)$, $(1000, 3000)$, $(3000, 1000)$, and $(3000, 3000)$. The remaining nine production wells are located at $(1, 1)$, $(1, 2000)$, $(1, 4000)$, $(2000, 1)$, $(2000, 2000)$, $(2000, 4000)$, $(4000, 1)$, $(4000, 2000)$, and $(4000, 4000)$. 
To introduce additional complexity and observe the characteristics of tested methods under different scenarios, the injection wells operate using a gas-water alternating injection strategy. 
Gas is injected for the first 500 days at a rate of 4000 Mscf/day, followed by water injection for the next 500 days at a rate of 4000 stb/day. This cycle is repeated twice, resulting in a total simulation time of 2000 days.

Regarding the choice of solution methods, we use the fully implicit method as the basic solution strategy. The following notations are used for the different methods:
\begin{enumerate}
    \item \textbf{FIM}: global fully implicit method, as shown in Figure~\ref{fig:OCP-parallel01}.
    \item \textbf{CDDM-FIM}: using classical DDM-based FIM to provide initial solutions for FIM, where each subdomain is solved independently to a specified accuracy.
    \item \textbf{ADDM-FIM}: using adaptively coupled DDM-based FIM to provide initial solutions for FIM. The key difference from CDDM-FIM is that at the beginning of each time step, subdomains are adaptively coupled for solving, as shown in Figure~\ref{fig:OCP-parallel07}.
\end{enumerate}

\captionsetup[subfigure]{skip=0pt}
\begin{figure}[H] 
    \centering
    \begin{subfigure}{0.2\textwidth}
        \centering
        \includegraphics[width=\linewidth]{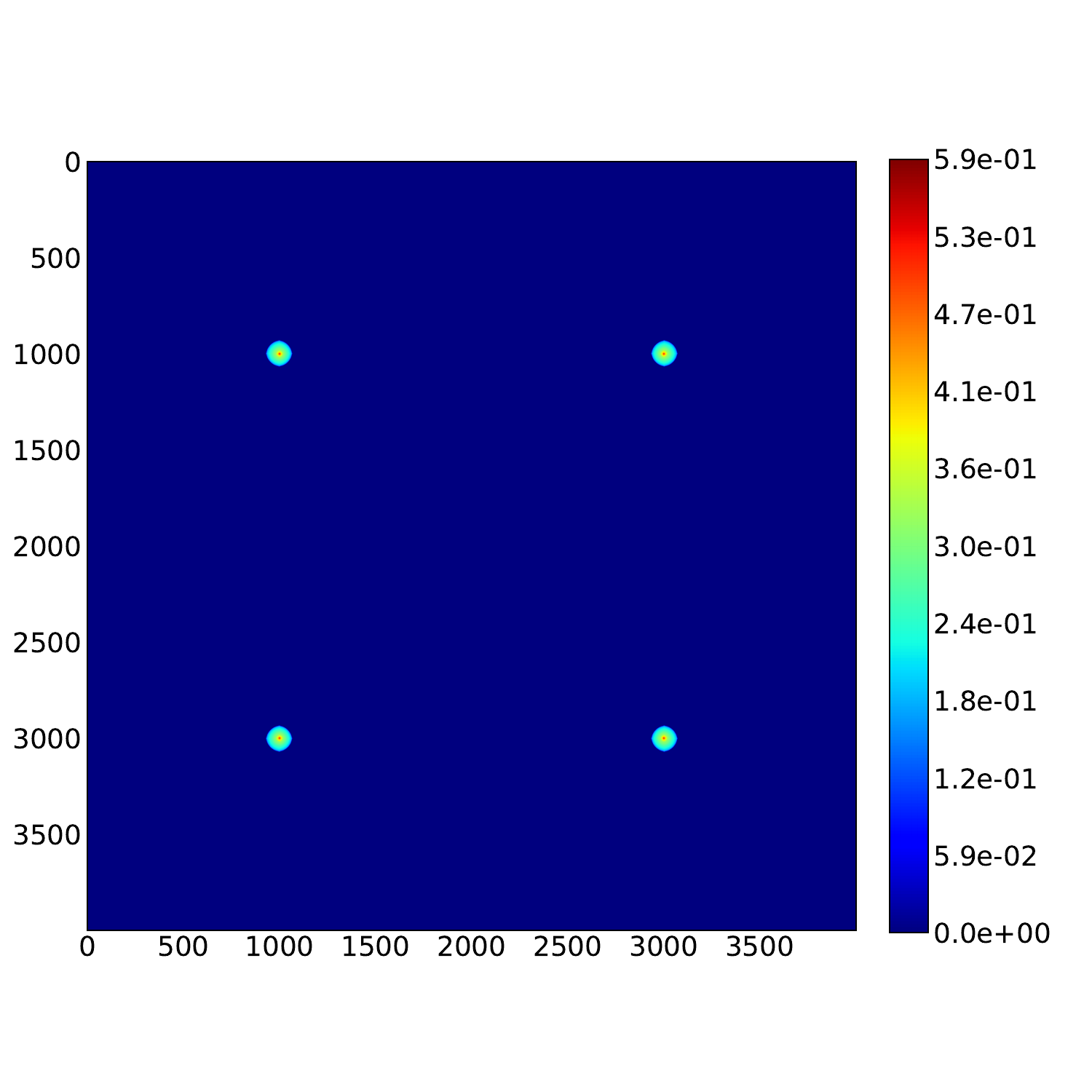}
        \caption{}
        \label{fig:case1:500day:sgas}
    \end{subfigure}
    \hspace{-0.02\textwidth}
    \begin{subfigure}{0.2\textwidth}
        \centering
        \includegraphics[width=\linewidth]{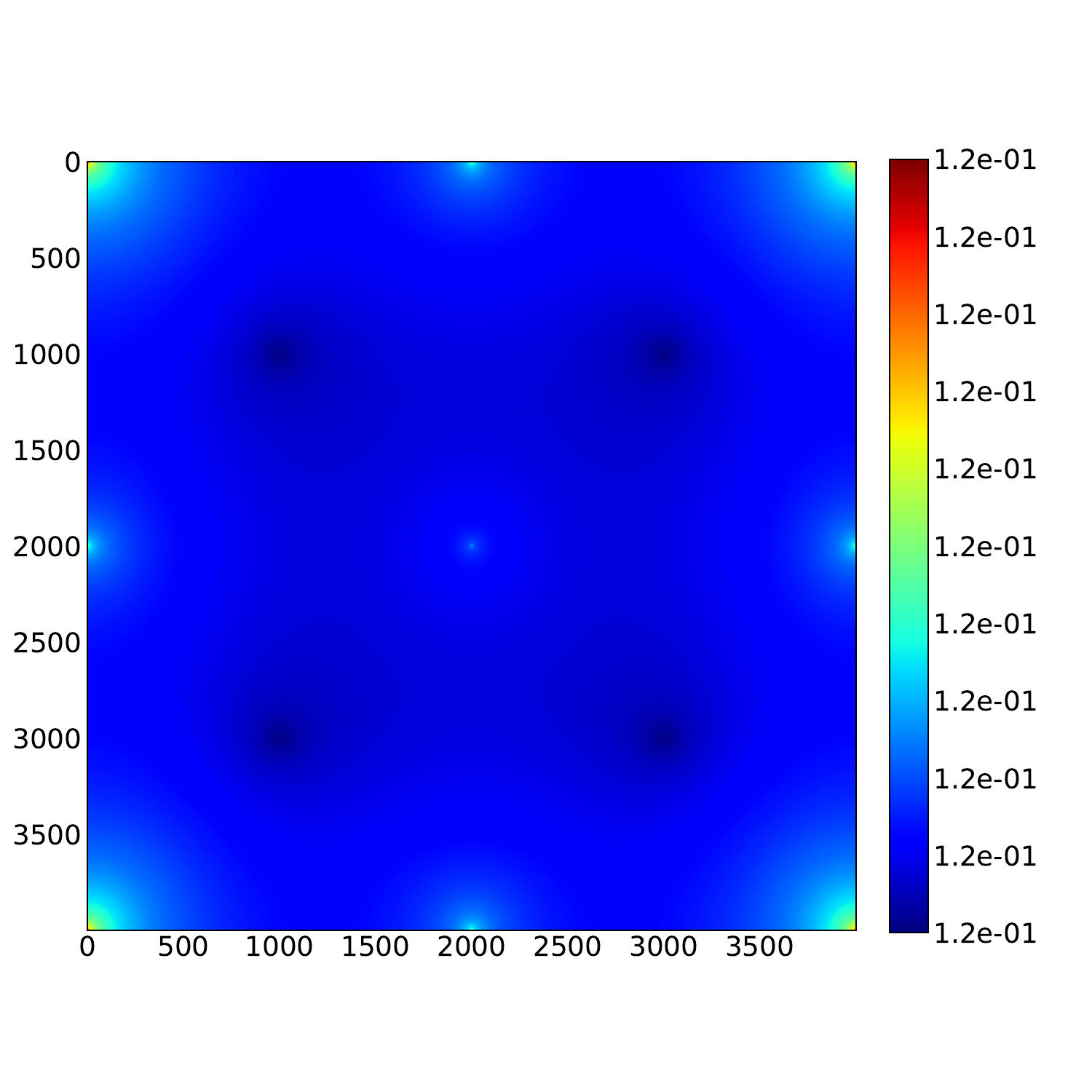}
        \caption{}
        \label{fig:case1:500day:swat}
    \end{subfigure}
    \hspace{-0.02\textwidth}
    \begin{subfigure}{0.2\textwidth}
        \centering
        \includegraphics[width=\linewidth]{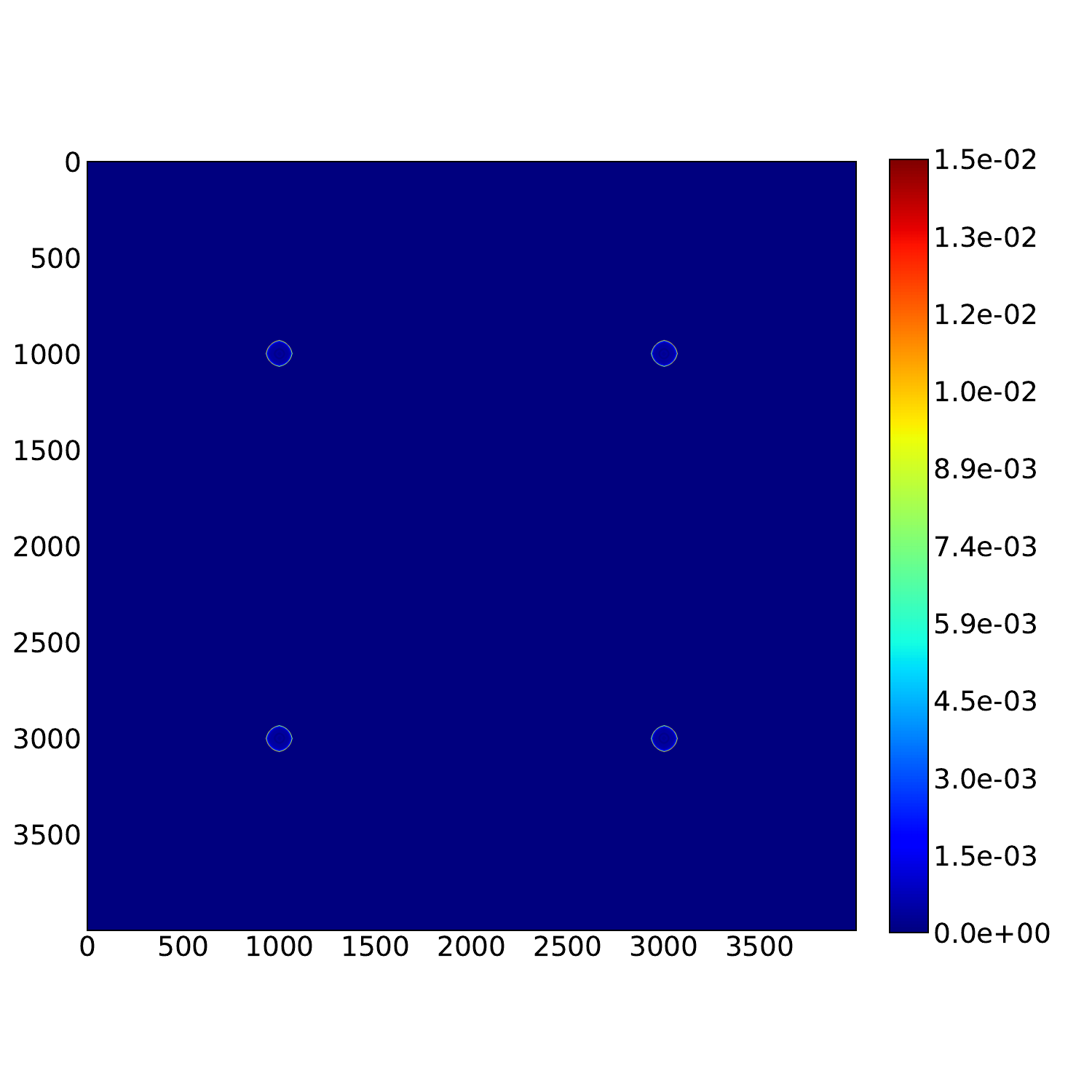}
        \caption{}
        \label{fig:case1:500day:dsgas}
    \end{subfigure}
    \hspace{-0.02\textwidth}
    \begin{subfigure}{0.2\textwidth}
        \centering
        \includegraphics[width=\linewidth]{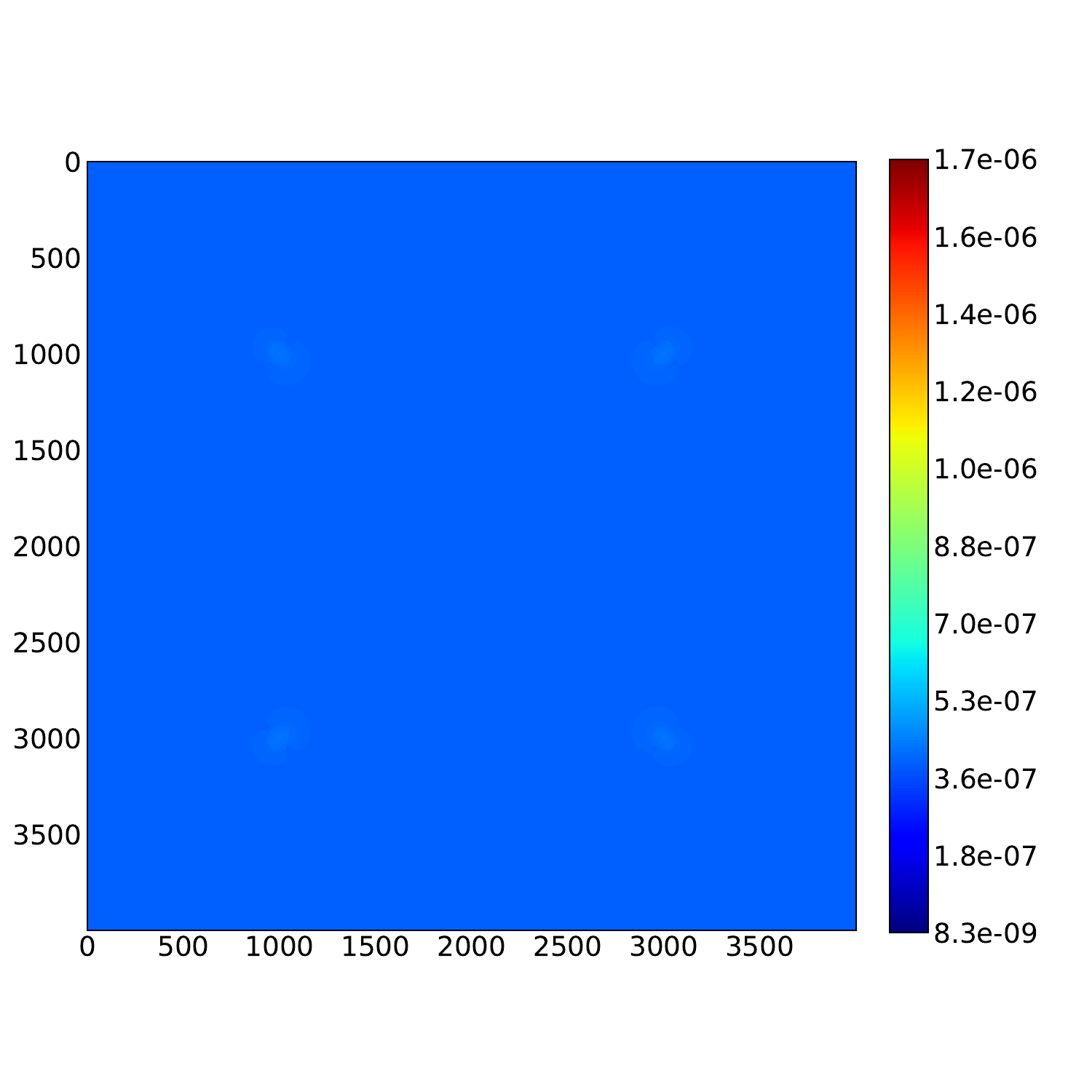}
        \caption{}
        \label{fig:case1:500day:dswat}
    \end{subfigure}
    \hspace{-0.02\textwidth}
    \begin{subfigure}{0.2\textwidth}
        \centering
        \includegraphics[width=\linewidth]{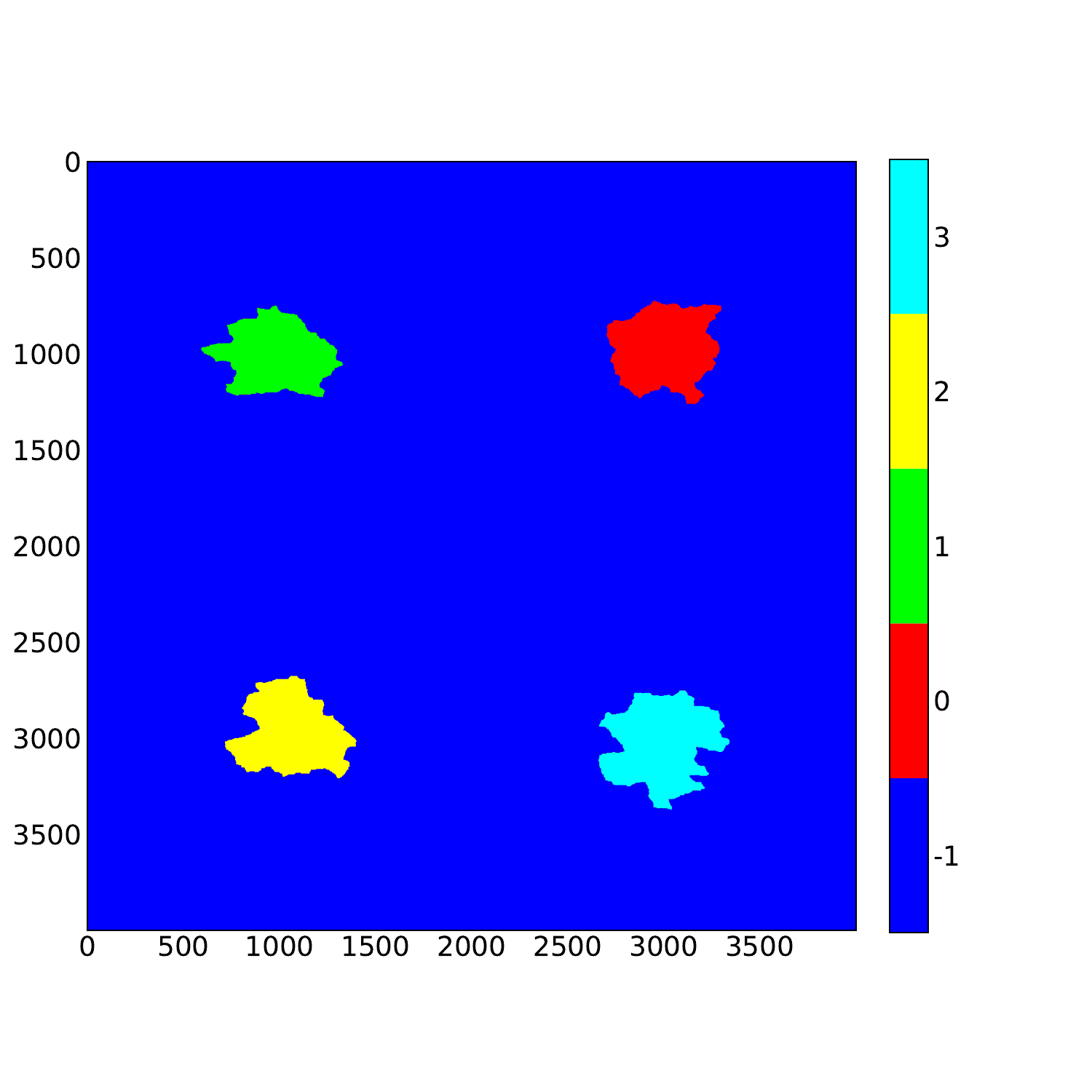}
        \caption{}
        \label{fig:case1:500day:csflag}
    \end{subfigure}

    \begin{subfigure}{0.2\textwidth}
        \centering
        \includegraphics[width=\linewidth]{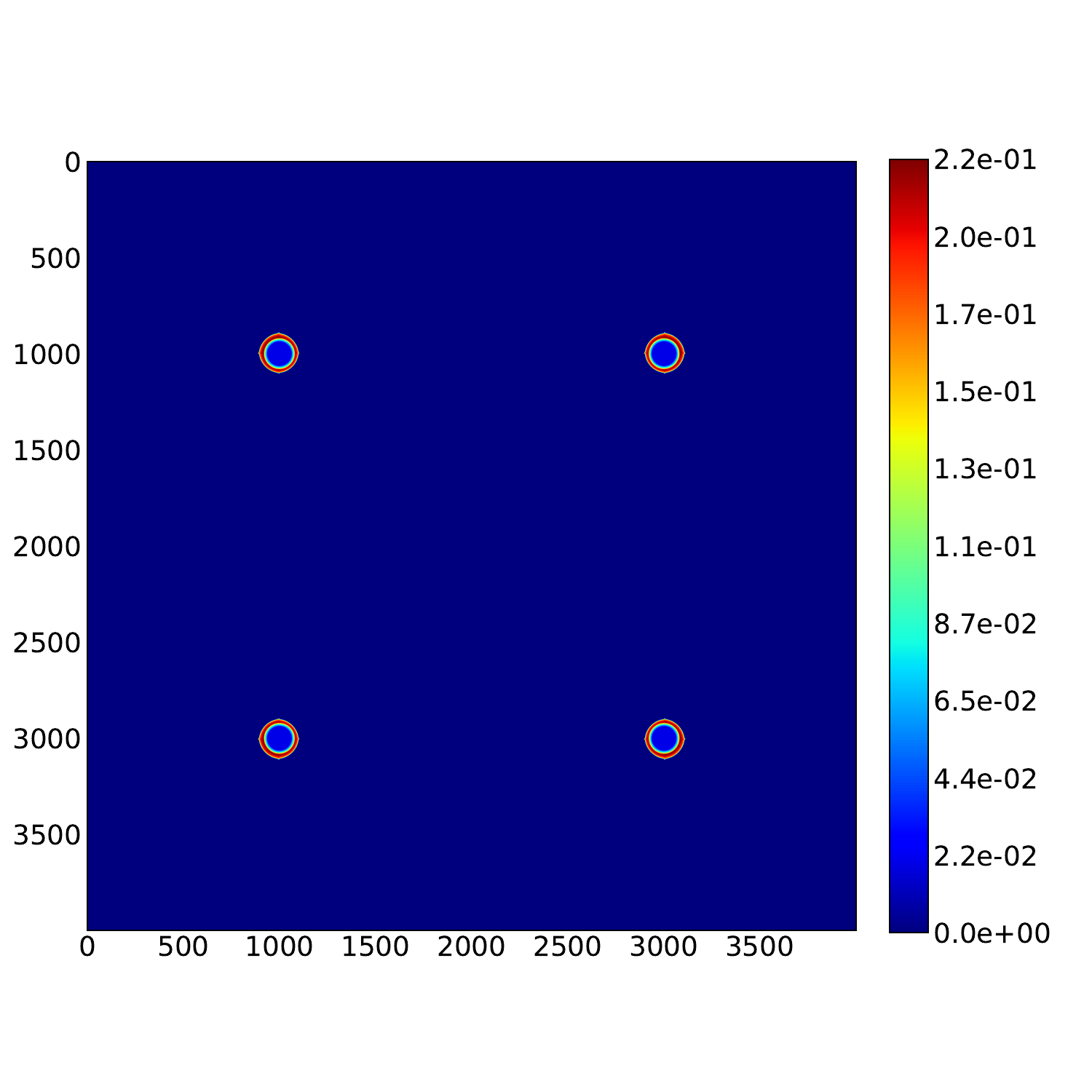}
        \caption{}
        \label{fig:case1:1000day:sgas}
    \end{subfigure}
    \hspace{-0.02\textwidth}
    \begin{subfigure}{0.2\textwidth}
        \centering
        \includegraphics[width=\linewidth]{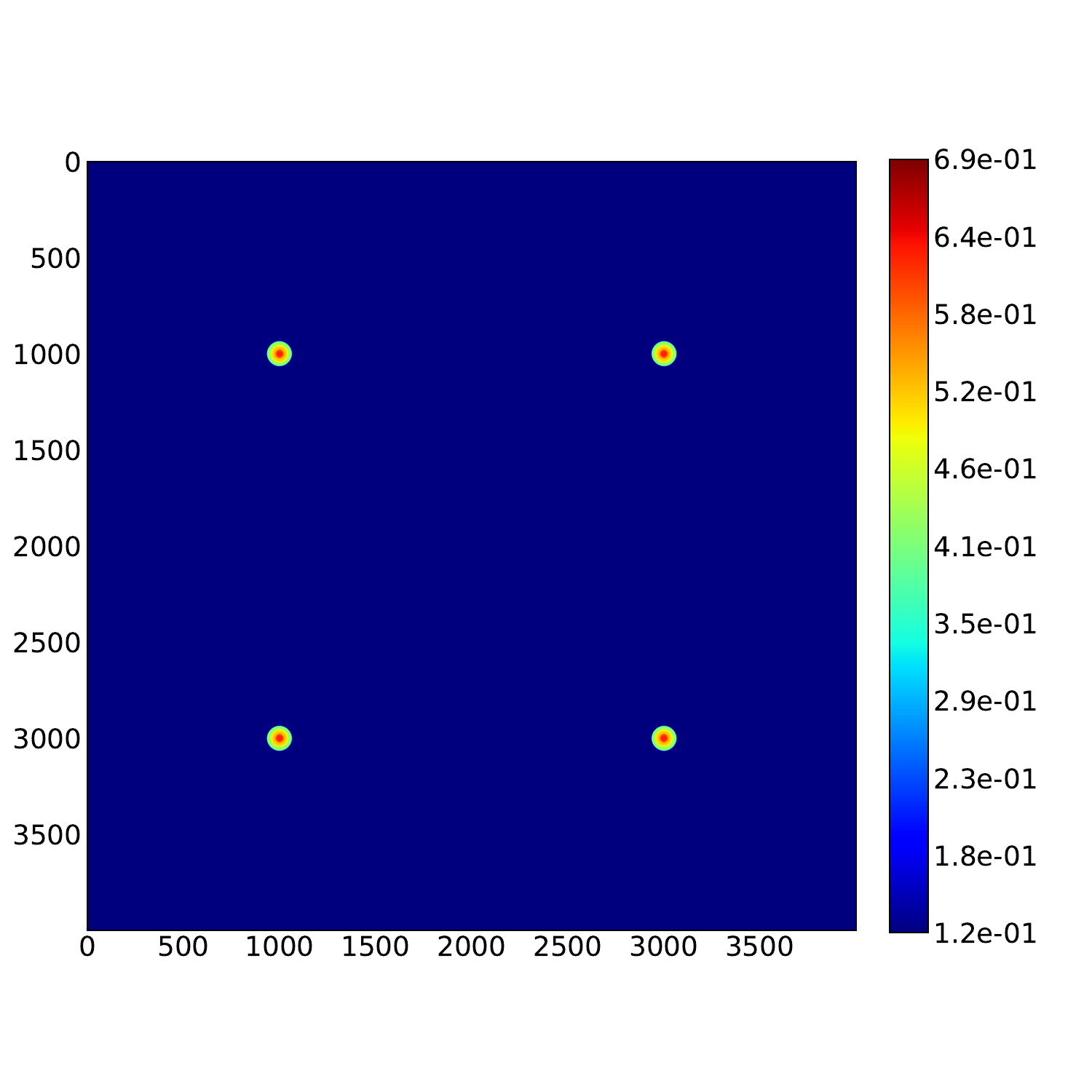}
        \caption{}
        \label{fig:case1:1000day:swat}
    \end{subfigure}
    \hspace{-0.02\textwidth}
    \begin{subfigure}{0.2\textwidth}
        \centering
        \includegraphics[width=\linewidth]{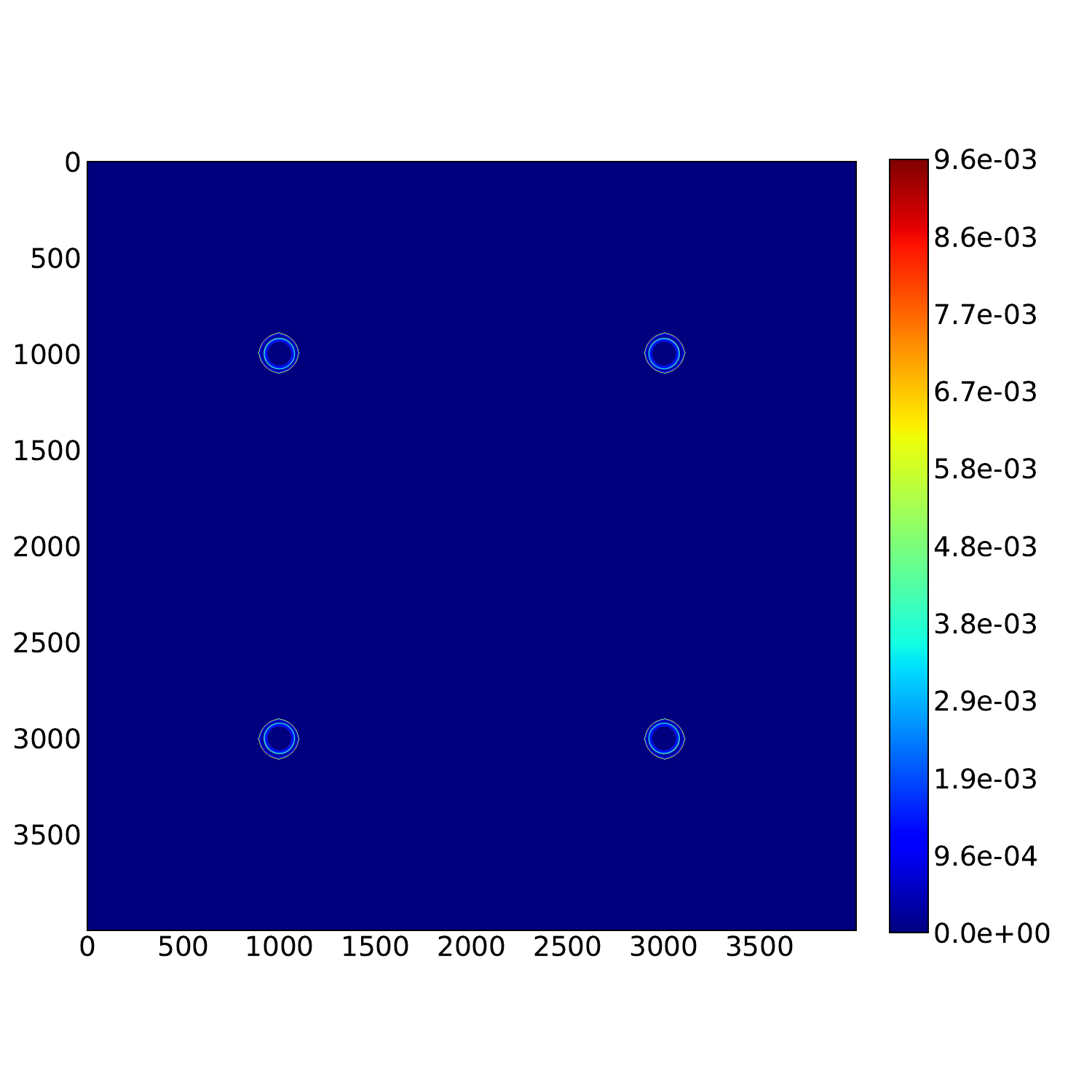}
        \caption{}
        \label{fig:case1:1000day:dsgas}
    \end{subfigure}
    \hspace{-0.02\textwidth}
    \begin{subfigure}{0.2\textwidth}
        \centering
        \includegraphics[width=\linewidth]{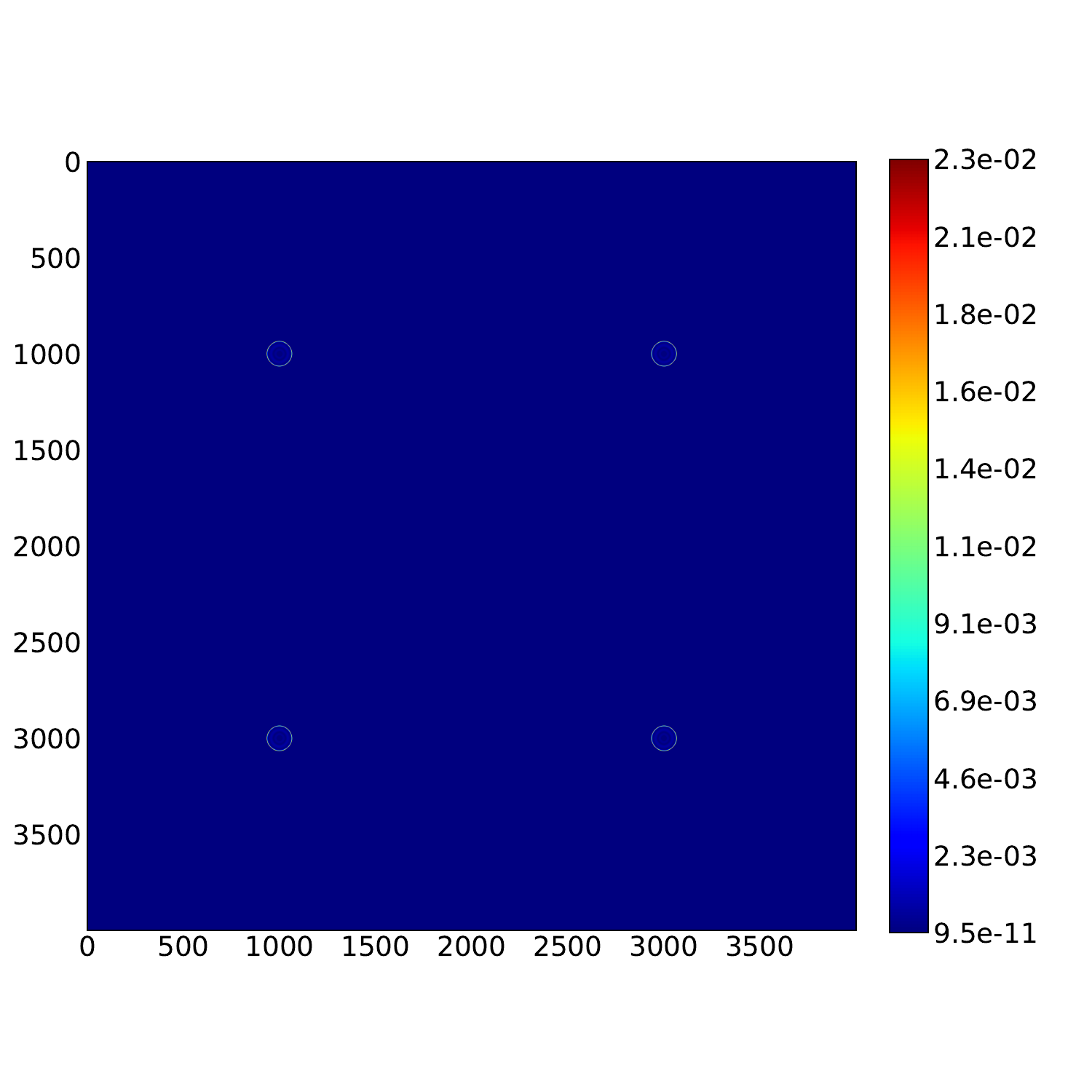}
        \caption{}
        \label{fig:case1:1000day:dswat}
    \end{subfigure}
    \hspace{-0.02\textwidth}
    \begin{subfigure}{0.2\textwidth}
        \centering
        \includegraphics[width=\linewidth]{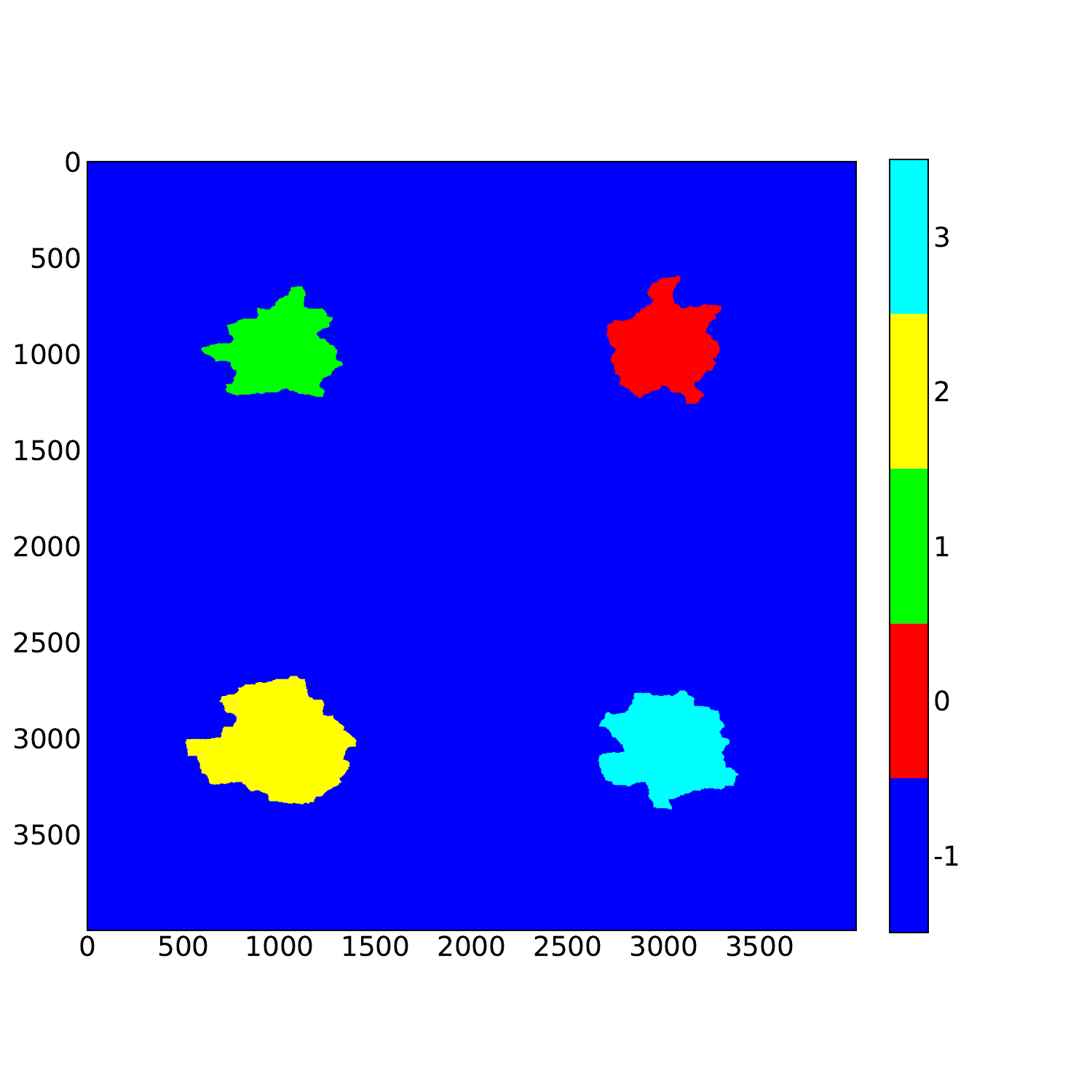}
        \caption{}
        \label{fig:case1:1000day:csflag}
    \end{subfigure}

    \begin{subfigure}{0.2\textwidth}
        \centering
        \includegraphics[width=\linewidth]{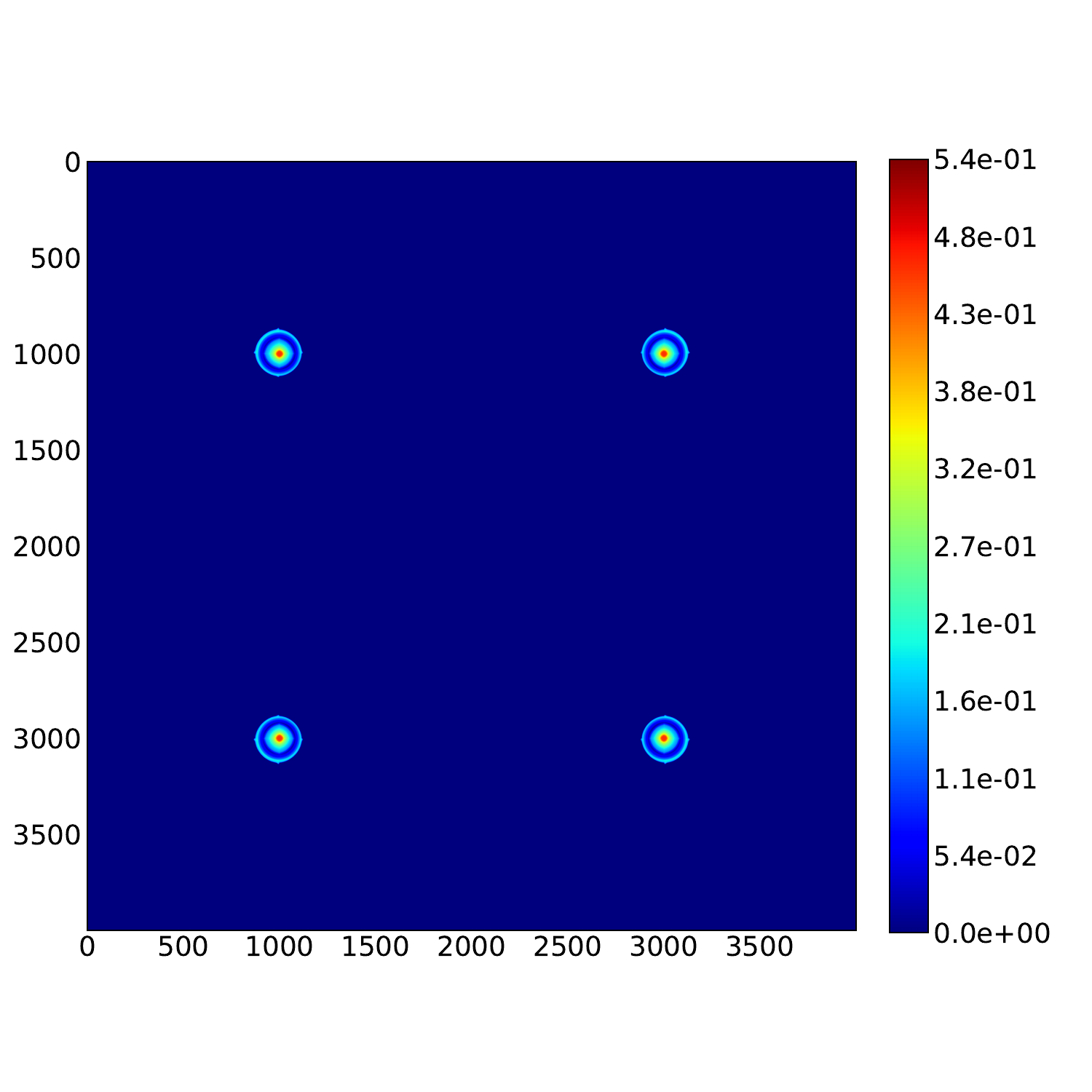}
        \caption{}
        \label{fig:case1:1500day:sgas}
    \end{subfigure}
    \hspace{-0.02\textwidth}
    \begin{subfigure}{0.2\textwidth}
        \centering
        \includegraphics[width=\linewidth]{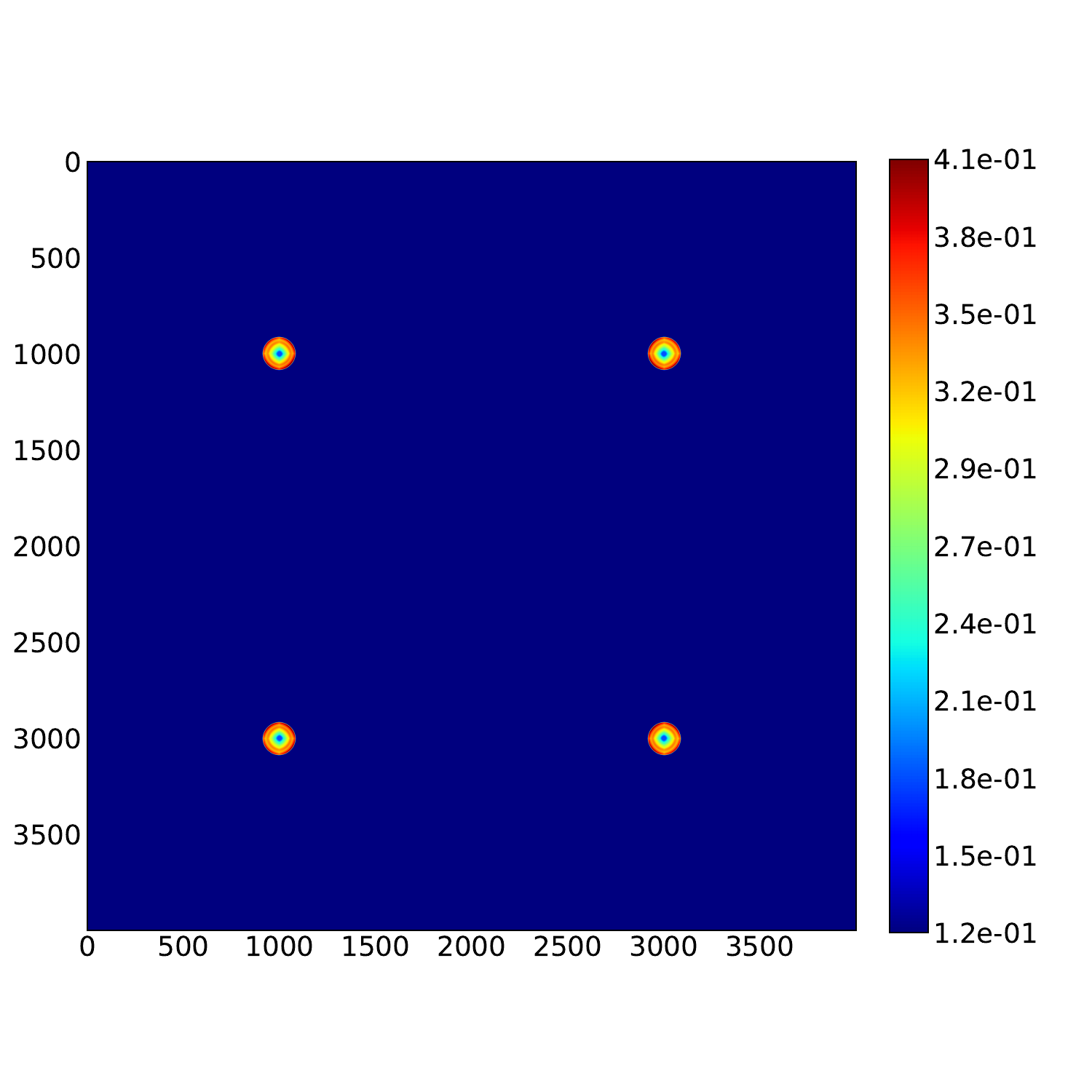}
        \caption{}
        \label{fig:case1:1500day:swat}
    \end{subfigure}
    \hspace{-0.02\textwidth}
    \begin{subfigure}{0.2\textwidth}
        \centering
        \includegraphics[width=\linewidth]{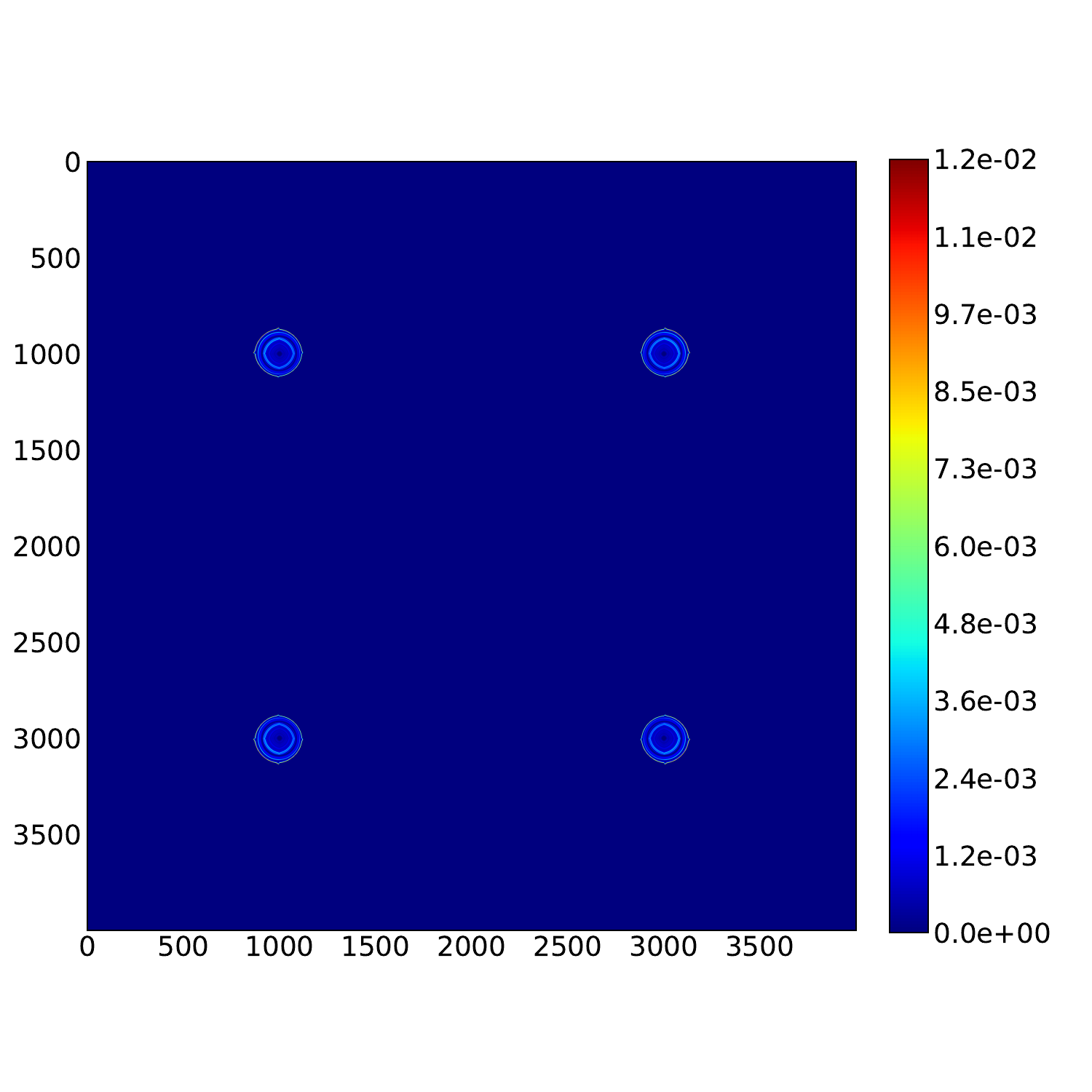}
        \caption{}
        \label{fig:case1:1500day:dsgas}
    \end{subfigure}
    \hspace{-0.02\textwidth}
    \begin{subfigure}{0.2\textwidth}
        \centering
        \includegraphics[width=\linewidth]{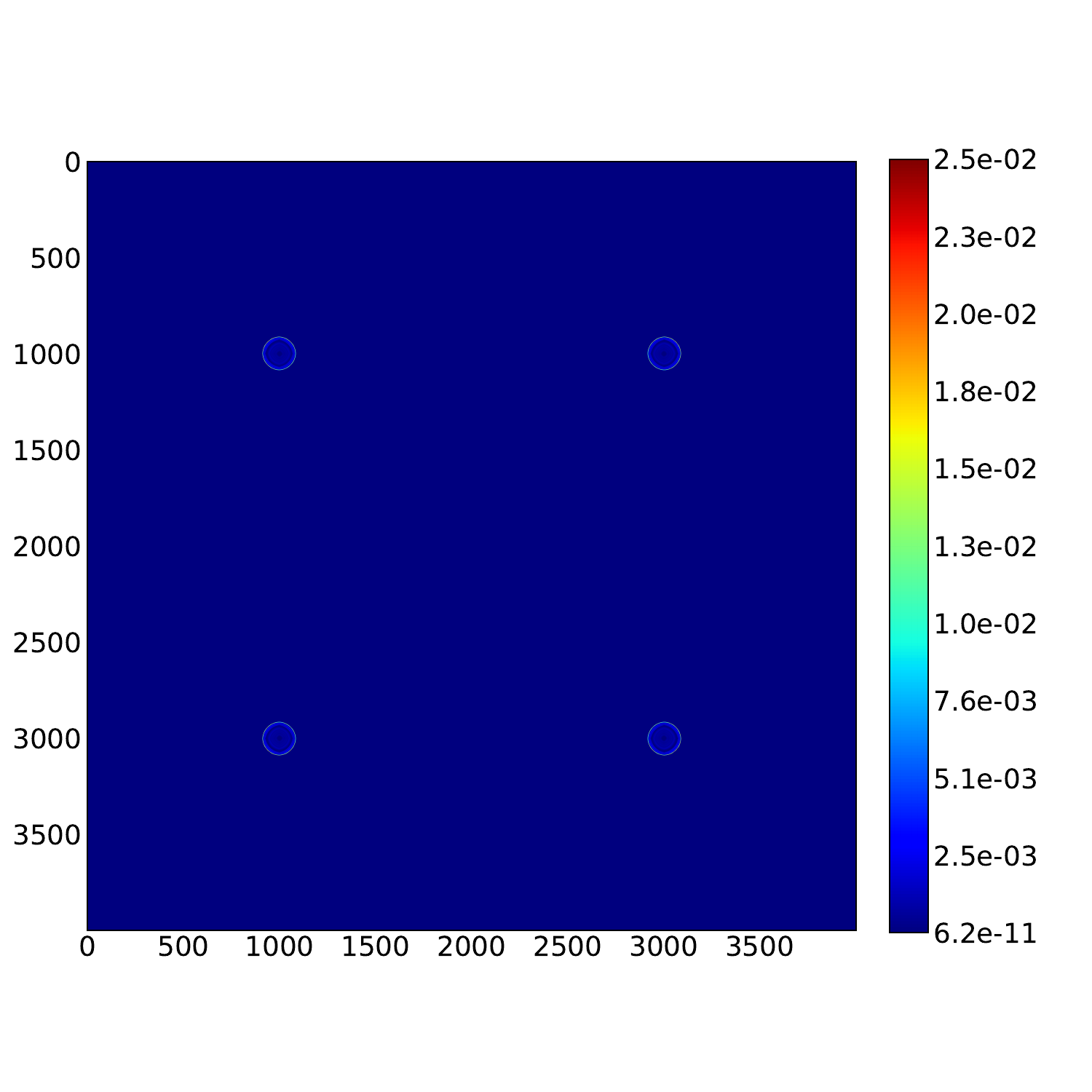}
        \caption{}
        \label{fig:case1:1500day:dswat}
    \end{subfigure}
    \hspace{-0.02\textwidth}
    \begin{subfigure}{0.2\textwidth}
        \centering
        \includegraphics[width=\linewidth]{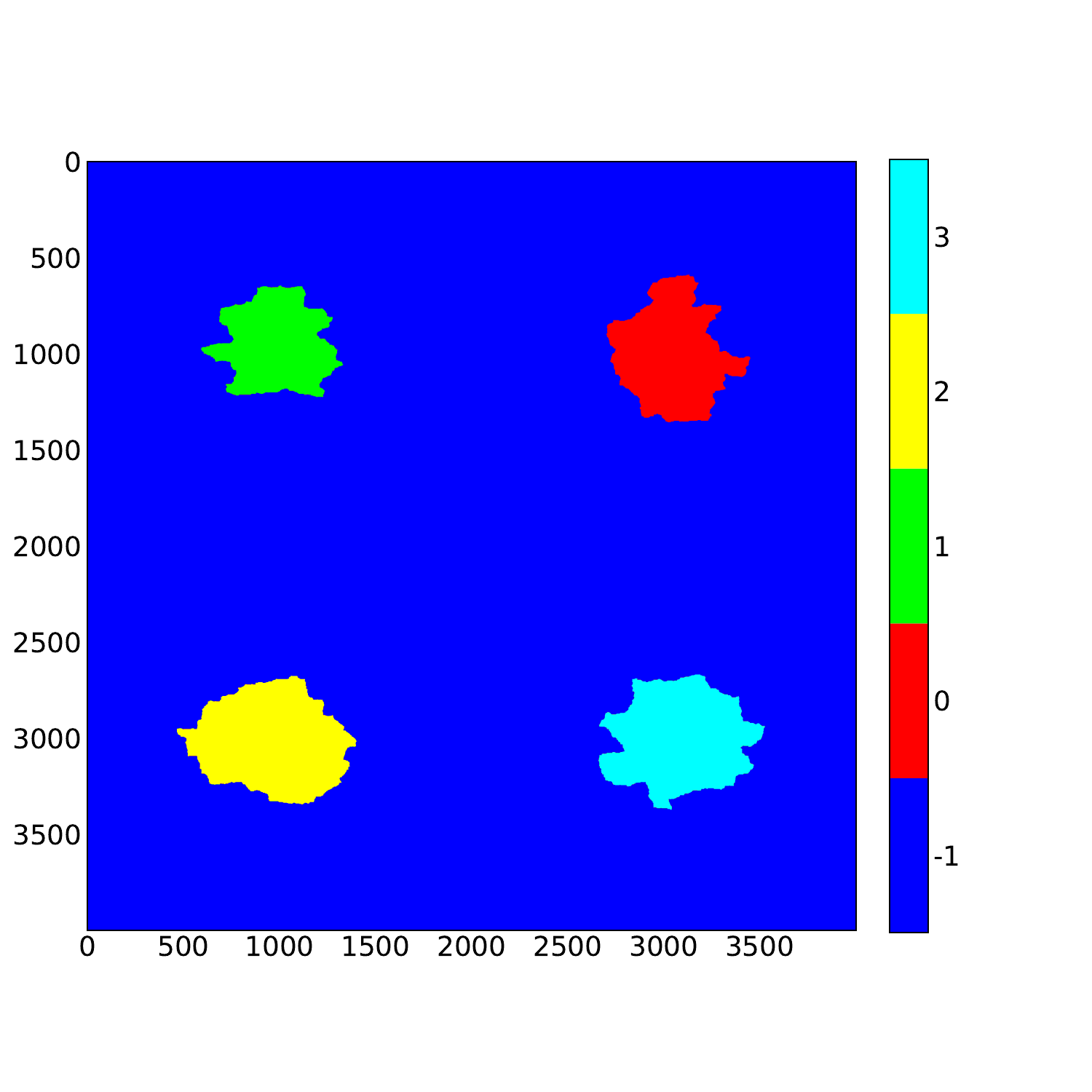}
        \caption{}
        \label{fig:case1:1500day:csflag}
    \end{subfigure}

    \begin{subfigure}{0.2\textwidth}
        \centering
        \includegraphics[width=\linewidth]{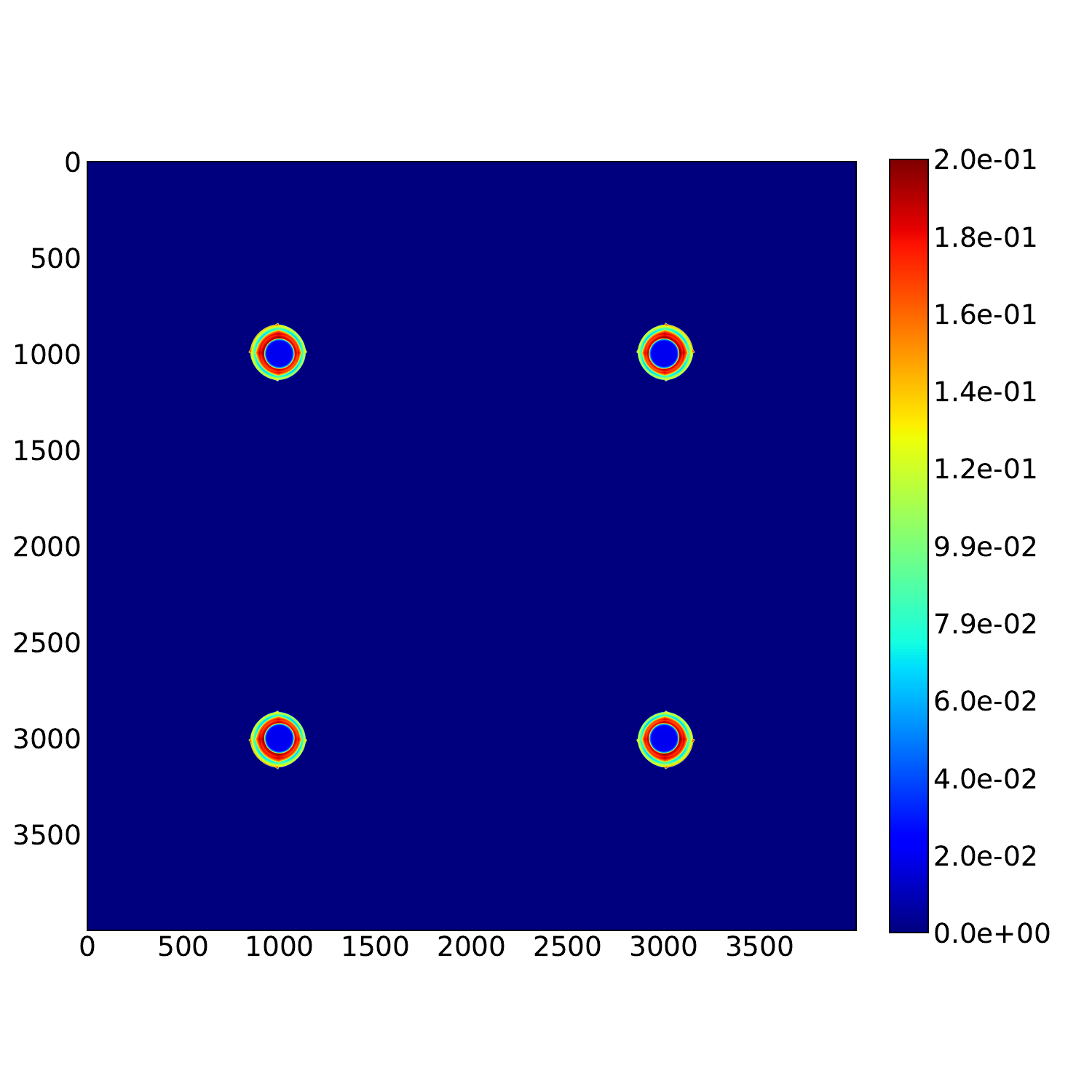}
        \caption{}
        \label{fig:case1:2000day:sgas}
    \end{subfigure}
    \hspace{-0.02\textwidth}
    \begin{subfigure}{0.2\textwidth}
        \centering
        \includegraphics[width=\linewidth]{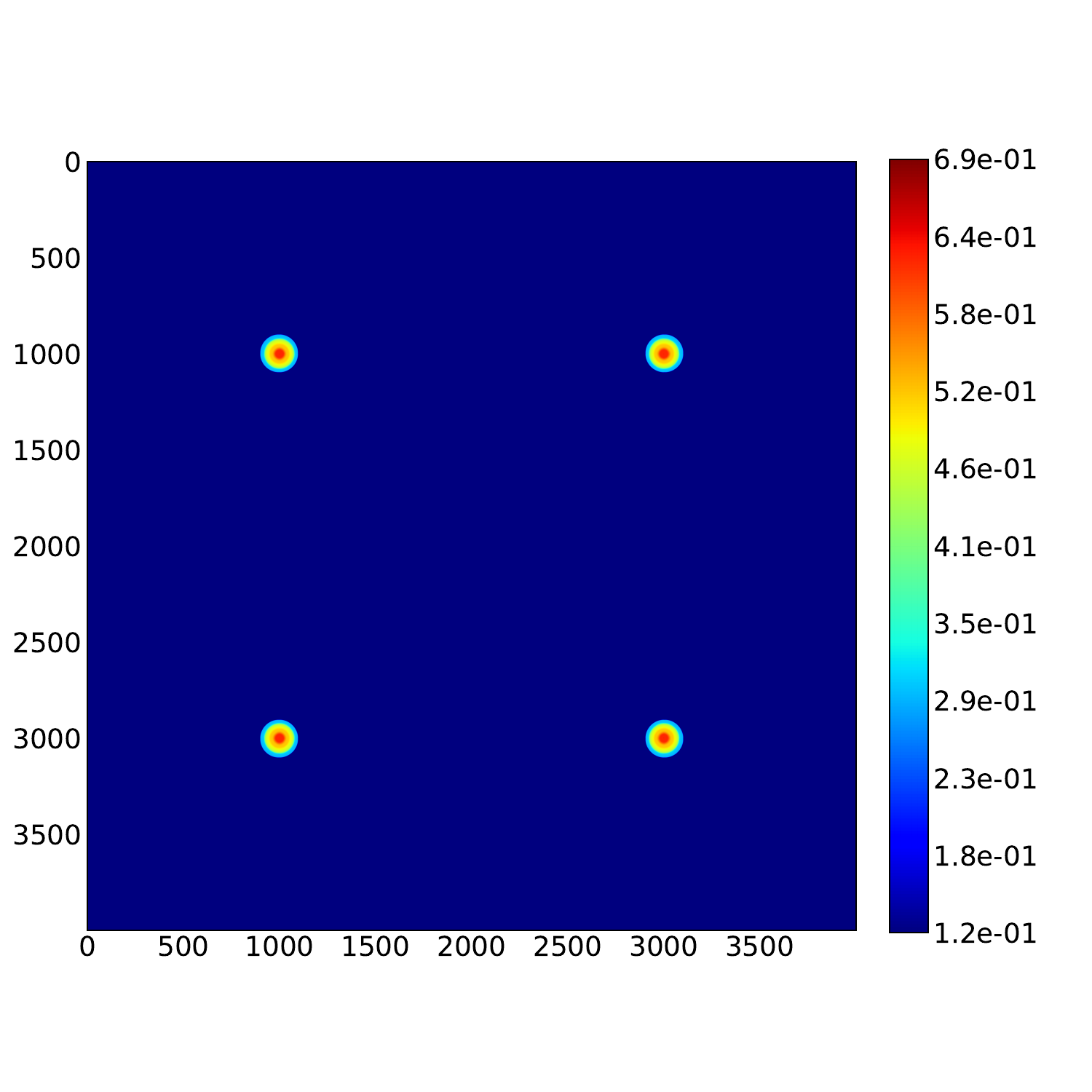}
        \caption{}
        \label{fig:case1:2000day:swat}
    \end{subfigure}
    \hspace{-0.02\textwidth}
    \begin{subfigure}{0.2\textwidth}
        \centering
        \includegraphics[width=\linewidth]{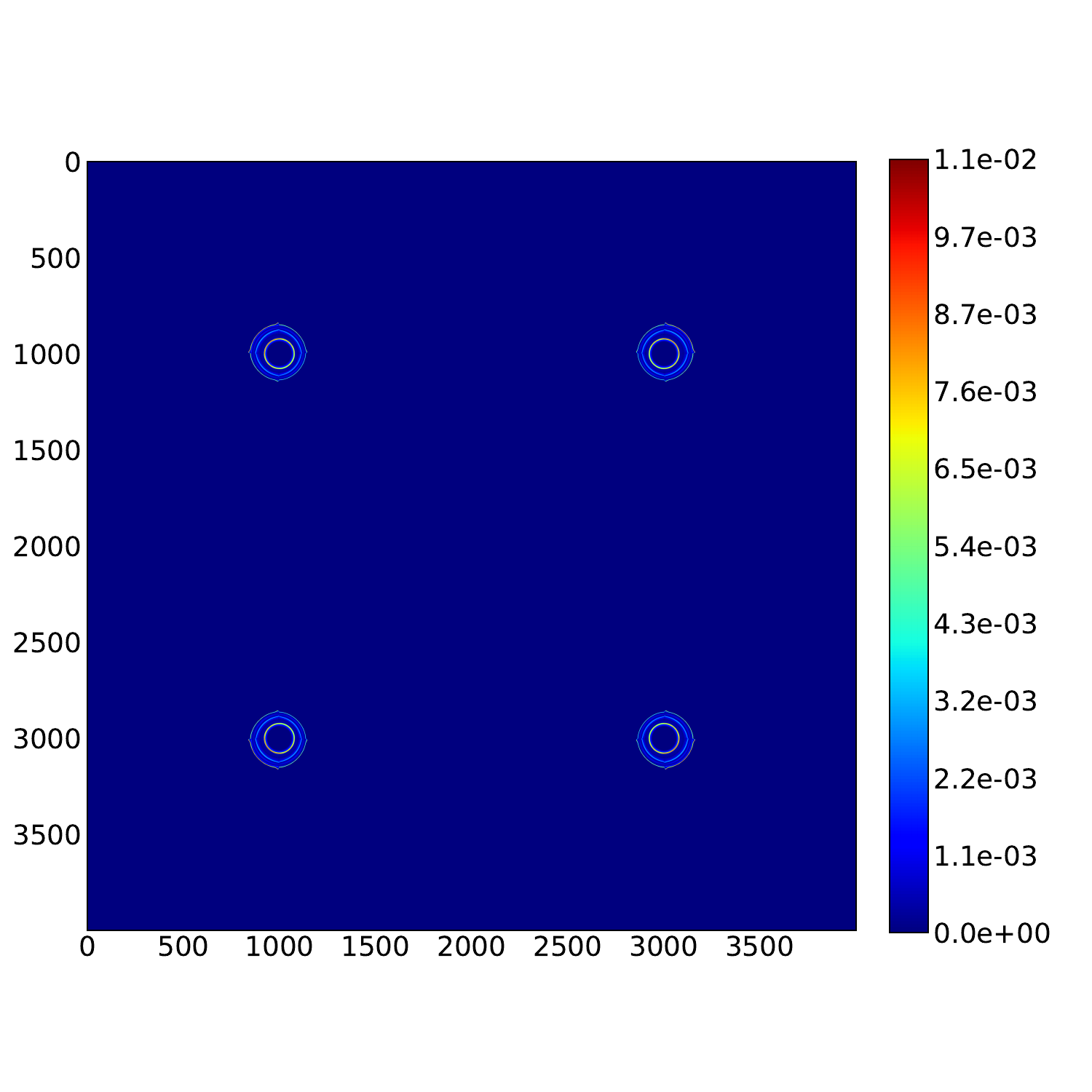}
        \caption{}
        \label{fig:case1:2000day:dsgas}
    \end{subfigure}
    \hspace{-0.02\textwidth}
    \begin{subfigure}{0.2\textwidth}
        \centering
        \includegraphics[width=\linewidth]{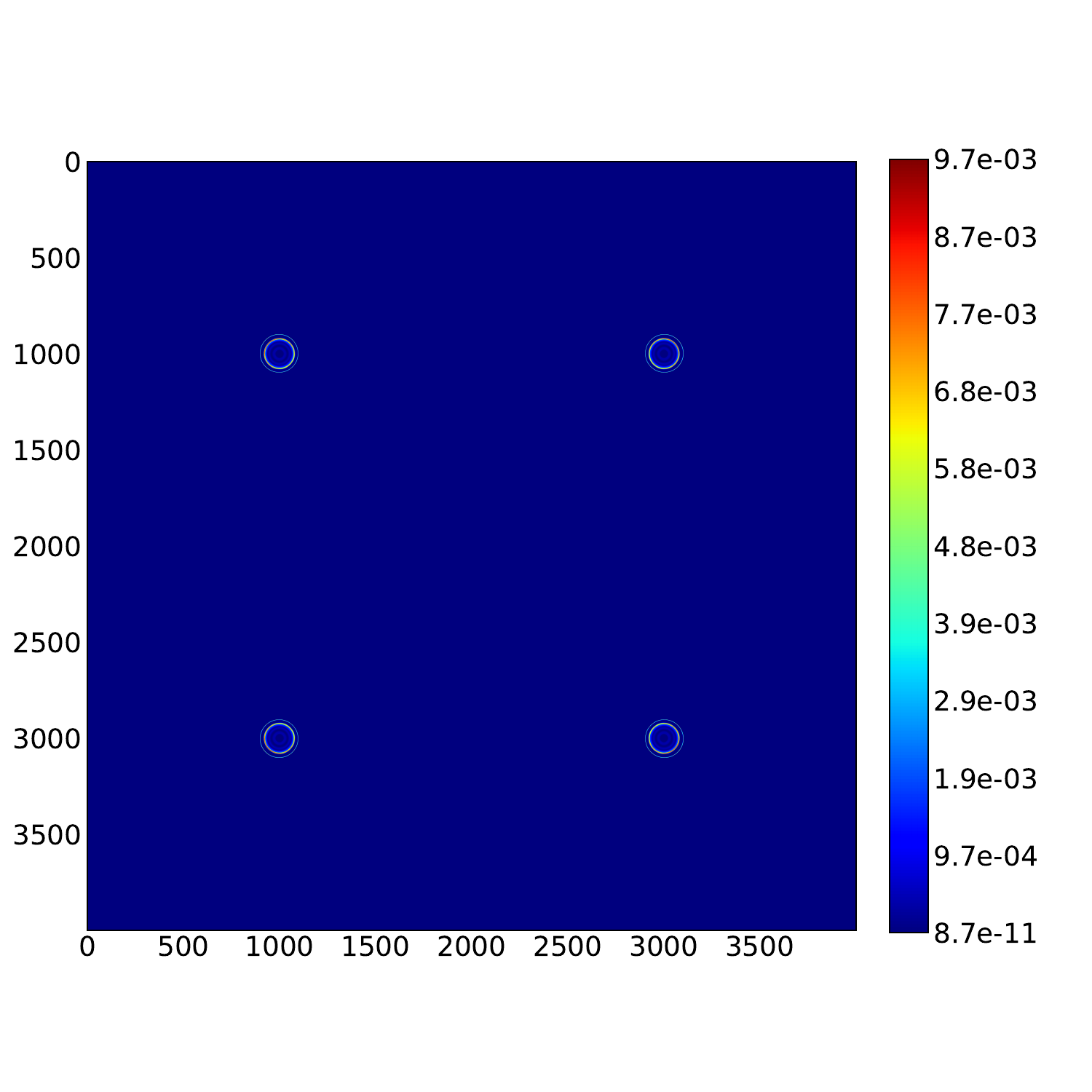}
        \caption{}
        \label{fig:case1:2000day:dswat}
    \end{subfigure}
    \hspace{-0.02\textwidth}
    \begin{subfigure}{0.2\textwidth}
        \centering
        \includegraphics[width=\linewidth]{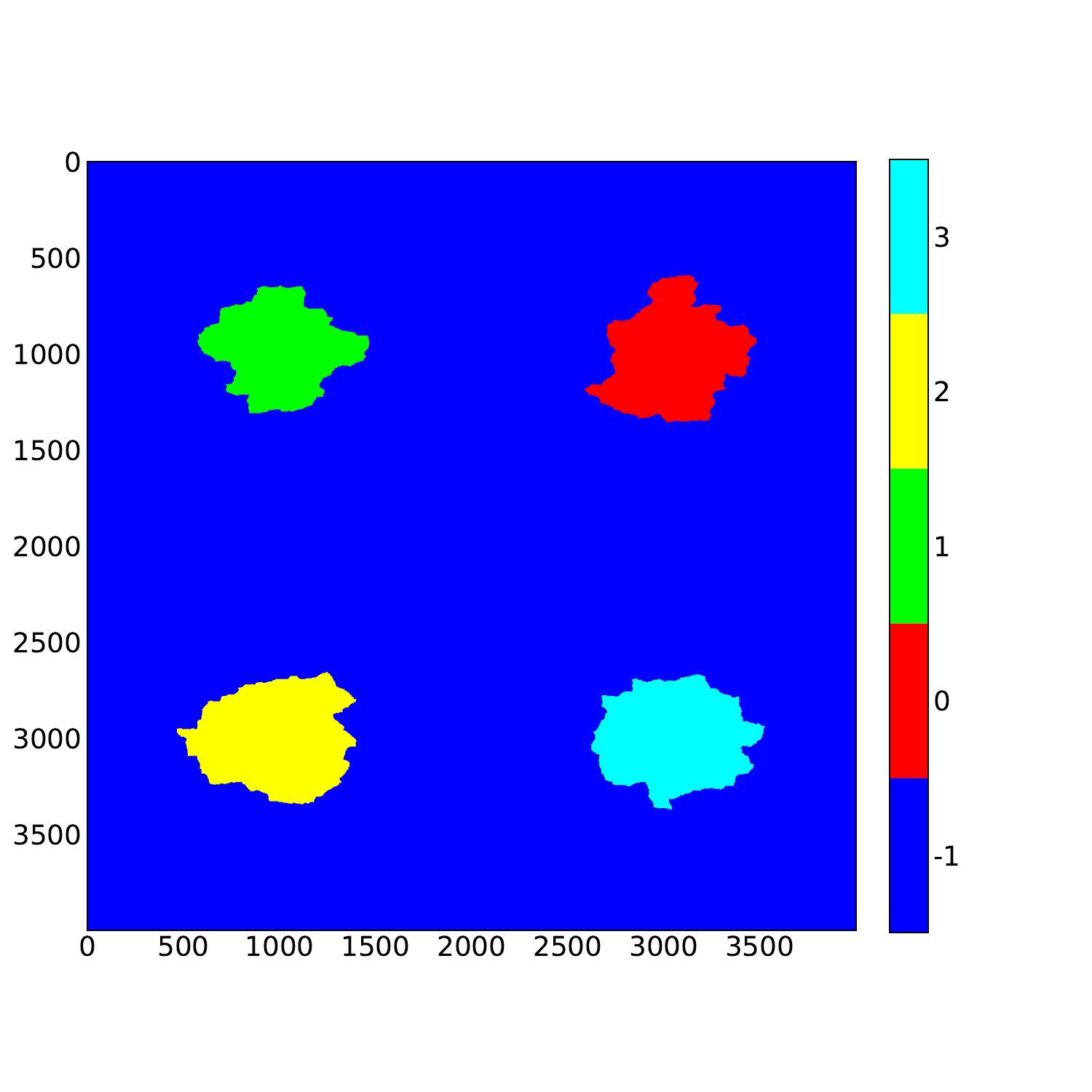}
        \caption{}
        \label{fig:case1:2000day:csflag}
    \end{subfigure}
    
    \caption{The four rows in the figure represent the simulation results at days 500, 1000, 1500, and 2000, respectively. The first four columns show the gas saturation, the water saturation, the change in gas saturation compared to the previous time step, and the change in water saturation compared to the previous time step, respectively. The last column shows the coupling of subdomains. Subdomains marked in blue are solved independently, while the remaining subdomains, if colored the same, are solved together.}
    \label{fig:case1:2D}
\end{figure}

For solving the linear systems, the CPR-type solvers are used to solve both global and local linear problems, as previously discussed.
The residual control for the global nonlinear and linear problems is set to $10^{-4}$, while for the local nonlinear and linear problems, it is set to $10^{-2}$.
We use 1000 processors for the computation.
Considering the total grid size of 16 million and 4 degrees of freedom per cell, each computing process is allocated an average of 16,000 grid cells, corresponding to 64,000 degrees of freedom. 

Figure~\ref{fig:case1:2D} presents key information about the reservoir state at days 500, 1000, 1500, and 2000, reflecting to some extent the philosophy of the subdomain adaptive coupling strategy.
For instance, during the first 500 days, as gas is continuously injected, the gas phase saturation near the injection wells increases, and the outer boundary of the gas phase expands. 
Over time, the flow further inside the gas phase moving interface gradually stabilizes, while the flow around the interface remains intense, indicating that the nearby coupling relationships are significant. 
Therefore, subdomains containing these regions should be solved in a coupled manner, as shown in Figure~\ref{fig:case1:500day:csflag}.
As water continues to be injected after the 500th day, new water phase moving interfaces form and move continuously. Simultaneously, the previously formed gas phase moving interfaces also expand. Consequently, to fully cover these moving interfaces, the number of coupled subdomains increases.
It should be noted that, due to the fast alternating injection of gas and water, the water phase moving interface and gas phase moving interface are not far apart, leading to a continuous increase in the area of the coupled regions. However, each coupled region remains relatively small compared to the total domain, as shown in the last column of Figure~\ref{fig:case1:2D}.

\begin{figure}[H] 
    \centering
    \begin{subfigure}{0.33\textwidth}
        \centering
        \includegraphics[width=\linewidth]{figures/results/case1/PARTITION.pdf}
        \caption{}
        \label{fig:cas1:partition}
    \end{subfigure} 
    \hspace{-0.02\textwidth}
    \begin{subfigure}{0.335\textwidth}
        \centering
        \includegraphics[width=\linewidth]{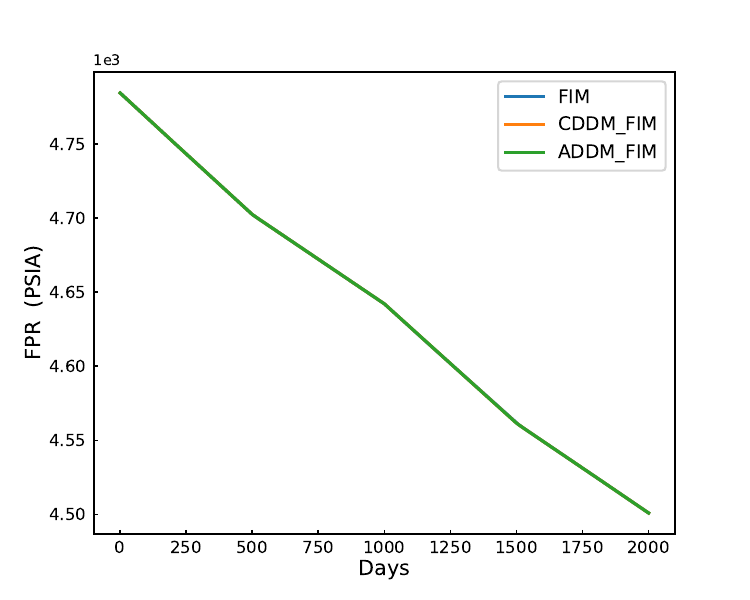}
        \caption{}
        \label{fig:cas1:FPR}
    \end{subfigure}
    \hspace{-0.03\textwidth}
    \begin{subfigure}{0.335\textwidth}
        \centering
        \includegraphics[width=\linewidth]{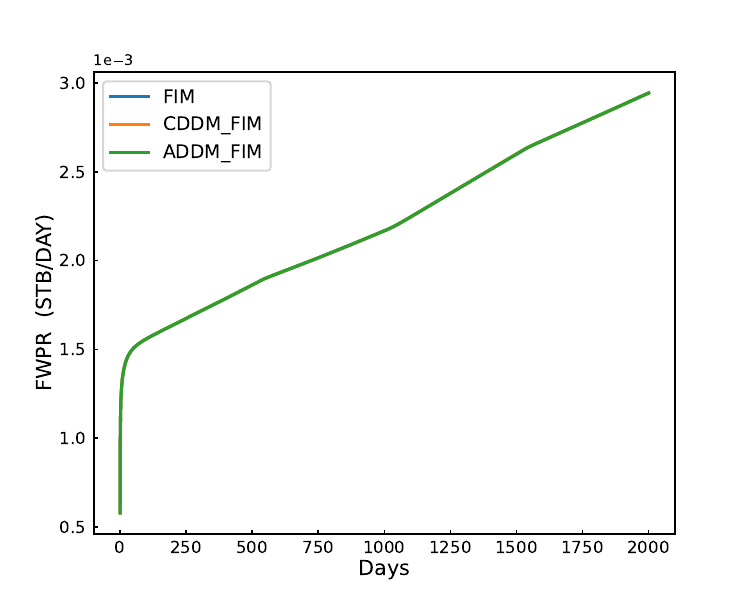}
        \caption{}
        \label{fig:cas1:FWPR}
    \end{subfigure}

    \begin{subfigure}{0.335\textwidth}
        \centering
        \includegraphics[width=\linewidth]{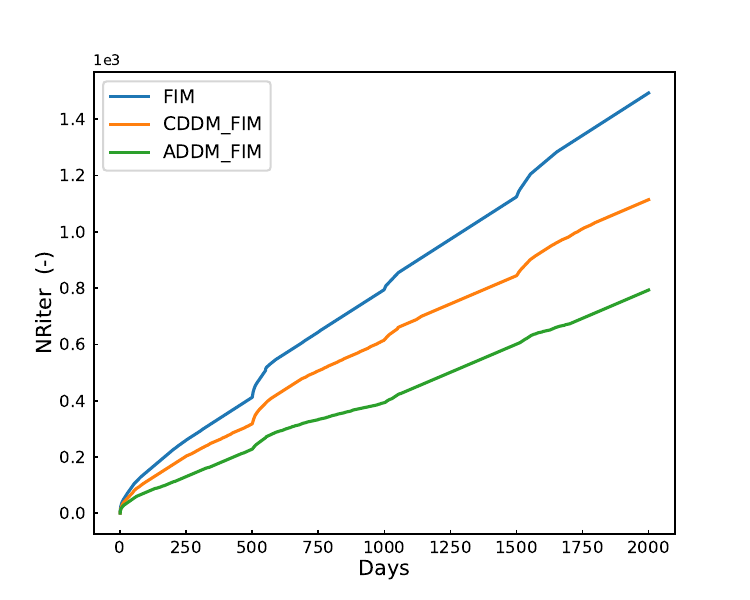}
        \caption{}
        \label{fig:cas1:NRiter}
    \end{subfigure}
    \hspace{-0.025\textwidth}
    \begin{subfigure}{0.335\textwidth}
        \centering
      \includegraphics[width=\linewidth]{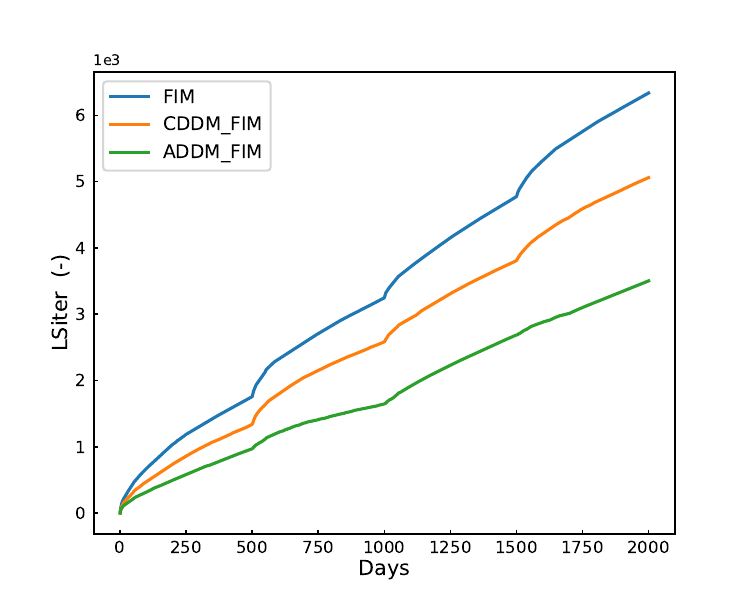}
        \caption{}
        \label{fig:cas1:LSiter}
    \end{subfigure}
    \hspace{-0.03\textwidth}
    \begin{subfigure}{0.335\textwidth}
        \centering
        \includegraphics[width=\linewidth]{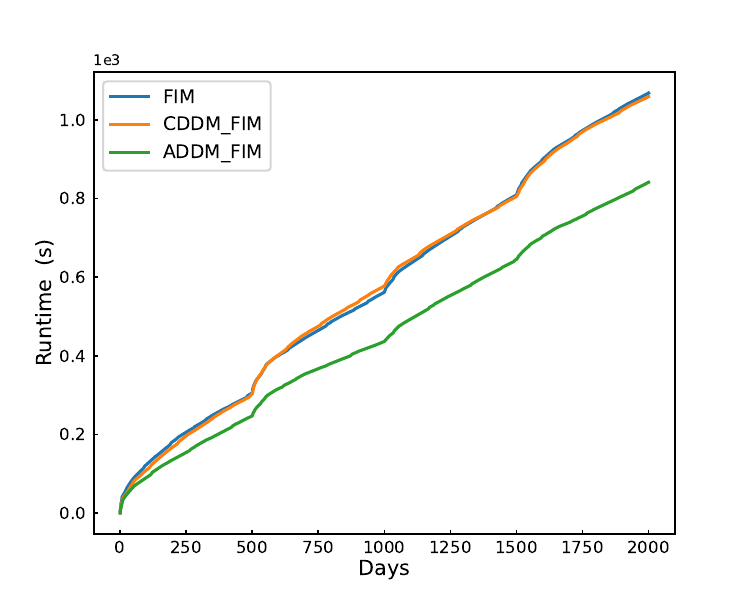}
        \caption{}
        \label{fig:cas1:Runtime}
    \end{subfigure}
    \caption{Figure (a) displays the domain partitioning by 1000 processes using ParMetis, with colors corresponding to the rank of each process. Figures (b) and (c) shows the field average pressure and field water production rate, respectively, which verify the consistency of the results obtained by the three methods. Figures (d), (e), and (f) illustrates the computational performance of the three methods from three perspectives: the cumulative global Newton-Raphson iterations, the cumulative global linear iterations, and the total runtime.}
    \label{fig:case1:res}
\end{figure}

Figure~\ref{fig:case1:res} compares the computational results of these three methods.
As expected, by providing suitable initial solutions, CDDM-FIM significantly reduces the number of required global Newton-Raphson iterations, which in turn greatly decreases the number of global linear iterations. 
Unfortunately, however, these reductions are not sufficient to offset the additional costs incurred in solving the local problems, resulting in a total time cost that does not provide an advantage over FIM.
By employing a subdomain adaptively coupled solution strategy, it is possible to better capture critical information at a low cost, allowing the solutions of local problems to more closely approximate the global problem. 
This strategy further reduces the required global Newton-Raphson iterations and linear iterations, significantly saving computational time.

More specific results are listed in Table~\ref{tab:case1:res}. 
As can be seen, compared to FIM, CDDM-FIM reduces the global Newton-Raphson iterations and hence global linear iterations by 25.7\% and 21.8\%, respectively, while the total wall time is reduced by only 0.84\%.
In contrast, ADDM-FIM reduces the global Newton-Raphson iterations and global linear iterations by 47.1\% and 45.9\%, respectively, achieving a 21.3\% reduction in time cost.
In addition, FIM wasted 7 global Newton-Raphson iterations and 132 global linear iterations. 
Similar situations occurred in CDDM-FIM and ADDM-FIM, but these were shifted to their local counterparts. 
Using local methods to provide initial solutions for global methods allows for identifying issues encountered during the solving process at a lower cost and making earlier adjustments, such as reducing the time step length.
Notably, in the latter two methods, the average number of linear iterations required per Newton-Raphson iteration significantly increased. This phenomenon occurred in almost all the cases we tested, suggesting a potential drawback of these methods that requires further investigation.
On the other hand, it is foreseeable that as the problem size increases and more processors are used, the global problem-solving process will be limited by its parallel efficiency, whereas the cost of solving sub-problems will remain constant or even decrease. 
\begin{table}
  \centering
  \caption{Comparison of the number of time steps, global Newton-Raphson (NR) iterations, global linear solver (LS) iterations, average linear iterations per NR iteration, and runtime across the three methods. The numbers in parentheses indicate the wasted iterations due to non-convergence in nonlinear/linear solution or the emergence of non-physical results.}
  \label{tab:case1:res}
  \scalebox{0.9}{
  \begin{tabular}{cccccc}
    \toprule
    Method & NumTimeSteps & NRiters & LSiters & LS/NR & Wall Time (s) \\
    \midrule
    FIM & 455 & 1493 (7) & 6337 (132) & 4.24 & 1068 \\
    CDDM-FIM & 449 & 1114 (0) & 5061 (0) & 4.54 & 1059 \\
    ADDM-FIM & 449 & 793 (0) & 3503 (0) & 4.41 & 841 \\
  \bottomrule
\end{tabular}}
\end{table}

\section{Conclusion}\label{sec:Conclusion}
In this paper, we present OpenCAEPoro, an open-source numerical simulation software for solving multiphase and multicomponent flow problems in porous media. 
Based on general compositional model equations, we modularize and abstract the computations of various physical objects and processes, enhancing their reusability and extensibility. 
Notably, we have develop a unified and flexible framework for various solution methods. 
This framework supports dynamic switching between solution methods in a nonlinear iteration or during time marching. 
Furthermore, we have developed an adaptively coupled domain decomposition method, which allows subdomains to be solved in an adaptively coupled manner. 
Our experiments show that using this method to provide initial solutions for the global method can significantly accelerate the simulation process.

\setlength{\nomlabelwidth}{2cm}
\setlength{\nomitemsep}{0.2em}
\nomenclature[A, 01]{$i$}{index of component}
\nomenclature[A, 02]{$j$}{index of phase}
\nomenclature[A, 03]{$t$}{time}
\nomenclature[A, 04]{$g$}{gravitational acceleration constant}
\nomenclature[A, 05]{$z$}{depth}
\nomenclature[A, 06]{$T$}{temperature}
\nomenclature[B, 01]{$n_{c}$}{number of components}
\nomenclature[B, 02]{$n_{p}$}{number of phases}
\nomenclature[C, 01]{$N_{i}$}{molar concentration of component $i$}
\nomenclature[C, 02]{$Q_{i}$}{volumetric molar injection or production rate for component $i$}
\nomenclature[D, 01]{$V_{f}$}{fluid volume}
\nomenclature[D, 02]{$P_{j}$}{pressure of phase $j$}
\nomenclature[D, 03]{$S_{j}$}{saturation of phase $j$}
\nomenclature[D, 04]{$\xi_{j}$}{molar density of phase $j$}
\nomenclature[D, 05]{$\rho_{j}$}{mass density of phase $j$}
\nomenclature[D, 06]{$\mu_{j}$}{mass density of phase $j$}
\nomenclature[D, 07]{$\kappa_{rj}$}{relative permeability of phase $j$}
\nomenclature[D, 08]{$x_{ij}$}{mole fraction of component $i$ in phase $j$}
\nomenclature[D, 09]{$\mathbf{D}_{ij}$}{diffusion coefficient tensor of component $i$ in phase $j$}
\nomenclature[D, 10]{$\mathbf{u}_{j}$}{volumetric flow rate of phase $j$}
\nomenclature[D, 11]{$U_{j}$}{molar internal energy of phase $j$}
\nomenclature[D, 12]{$H_{j}$}{molar enthalpy of phase $j$}
\nomenclature[D, 13]{$q_{H,j}$}{volumetric enthalpy injection or production rate for phase $j$}
\nomenclature[D, 13]{$q_{\text{loss}}$}{volumetric heat loss to the overburden or underburden}
\nomenclature[F, 01]{$V_{p}$}{pore volume}
\nomenclature[F, 02]{$\phi$}{porosity of rock}
\nomenclature[F, 03]{$\boldsymbol{\kappa}$}{effective permeability of rock}
\nomenclature[F, 04]{$\boldsymbol{\kappa}_{T}$}{effective thermal conductivity of rock}
\nomenclature[F, 05]{$U_{r}$}{energy density of the rock}
\nomenclature[G, 01]{$\textit{flux\_v}_{j}$}{volume flow rate of phase $j$ within connection pairs}
\nomenclature[G, 02]{$\textit{flux\_N}_{i}$}{molar rate of components $i$ within connection pairs}
\printnomenclature



\section*{Acknowledgments}
This research is supported by the Strategic Priority Research Program of the Chinese Academy of Sciences, Grant No. XDB0640000.


\printbibliography


\renewcommand\theequation{\Alph{section}\arabic{equation}} 
\counterwithin*{equation}{section} 
\renewcommand\thefigure{\Alph{section}\arabic{figure}} 
\counterwithin*{figure}{section} 
\renewcommand\thetable{\Alph{section}\arabic{table}} 
\counterwithin*{table}{section} 

\end{document}